\documentclass[11pt]{article}
\newcommand{\rev}[1]{#1}
\baselineskip
\pdfoutput=1
\usepackage{amsmath}
\usepackage{amssymb}
\usepackage{graphicx}
\usepackage{color}
\usepackage{cite}
\usepackage{collref}
\usepackage{tabls}
\usepackage{hyperref}
\usepackage{multirow}
\DeclareMathOperator{\SU}{\mathrm{SU}}
\DeclareMathOperator{\Tr}{\mathrm{Tr}}

\newcommand{\Z}{\mathbb{Z}}
\newcommand{\redchisq}{\chi^2_{\tiny\mbox{red}}}

\newcommand{\SW}{S_{\mbox{\tiny{W}}}}
\newcommand{\Tc}{T_{\mbox{\tiny{c}}}}
\newcommand{\nconf}{n_{\mbox{\tiny{conf}}}}
\newcommand{\gammaE}{\gamma_{\mbox{\tiny{E}}}}
\newcommand{\betac}{\beta_{\mbox{\tiny{c}}}}

\begin{document}

\begin{titlepage}
\begin{center}
{\Large\bf Fine corrections in the effective string describing $\SU(2)$ Yang-Mills theory in three dimensions}
\end{center}
\vskip10mm
\centerline{F.~Caristo,$^{1}$ M.~Caselle,$^{1,2}$ N.~Magnoli,$^{3}$ A.~Nada,$^{1}$ M.~Panero,$^{1,2}$ and A.~Smecca$^{1}$}
\vskip10mm
\centerline\emph{\noindent $^{1}$ Department of Physics, University of Turin \& INFN, Turin}
\centerline\emph{Via Pietro Giuria 1, I-10125 Turin, Italy}
\vskip2mm
\centerline\emph{\noindent $^{2}$ Arnold-Regge Center, University of Turin}
\centerline\emph{Via Pietro Giuria 1, I-10125 Turin, Italy}
\vskip2mm
\centerline\emph{\noindent $^{3}$ Department of Physics, University of Genoa \& INFN, Genoa}
\centerline\emph{Via Dodecaneso 33, I-16146 Genoa, Italy}
\vskip10mm

\begin{abstract}
We present a study of the effective string that describes the infrared dynamics of $\SU(2)$ Yang-Mills theory in three dimensions. By combining high-precision lattice simulation results for Polyakov-loop correlators at finite temperatures close to (and less than) the deconfinement one with the analytical constraints from renormalization-group arguments, from the exact integrability of the two-dimensional Ising model that describes the universality class of the critical point of the theory, from conformal perturbation theory, and from Lorentz invariance, we derive tight quantitative bounds on the corrections to the effective string action beyond the Nambu-Got{\={o}} approximation. We show that these corrections are compatible with the predictions derived from a bootstrap analysis of the effective string theory, \rev{and have a value which does not allow to prove the  Axionic String \emph{Ansatz} for this model}.
\end{abstract}

\end{titlepage}

\section{Introduction}
\label{sec:introduction}

One of the most promising approaches to understand and model the non-perturbative behavior of confining Yang-Mills theories is the ``effective string theory'' (EST) description, in which the flux tube joining together a quark-antiquark pair is modeled as a thin vibrating string~\cite{Nambu:1974zg,Goto:1971ce,Luscher:1980ac,Luscher:1980fr,Polchinski:1991ax}. Recently, there has been a lot of progress in this context. In particular, it has been realized that the EST enjoys ``low-energy universality''~\cite{Meyer:2006qx,Luscher:2004ib, Aharony:2009gg, Aharony:2011gb, Gliozzi:2011hj, Gliozzi:2012cx, Aharony:2013ipa}: due to the peculiar features of the string action and to the symmetry constraints imposed by Poincar\'e invariance in the target space, the first few terms of the long-distance expansion of the string action are fixed and hence universal. This implies that the EST is much more predictive than typical effective theories, and in fact during the past few years its predictions have been be confirmed by many simulations in lattice gauge theories (for recent reviews, see for instance refs.~\cite{Aharony:2013ipa,Brandt:2016xsp,Caselle:2021eir}).

At the same time, it was also realized that the simplest Lorentz-invariant EST, which is the well-known Nambu-Got{\={o}} model~\cite{Nambu:1974zg,Goto:1971ce}, is an exactly integrable, irrelevant, perturbation of the two-dimensional free Gau{\ss}ian model~\cite{Dubovsky:2012sh}, driven by the $T\overline{T}$ operator of the $D-2$ free bosons\footnote{We will denote in the following with $D$ the number of spacetime dimensions of the target lattice theory and with $d\equiv D-1$ the number of spacelike directions.} that represent the transverse degrees of freedom of the string~\cite{Caselle:2013dra}. This observation stimulated much work and led to interesting novel results, whose relevance extends even beyond the original application to Yang-Mills theory~\cite{Dubovsky:2012wk, Dubovsky:2013gi, Dubovsky:2014fma, Cooper:2014noa, Dubovsky:2015zey, Conkey:2016qju, Dubovsky:2017cnj, Dubovsky:2018bmo}. In particular, these findings are at the basis of an S-matrix bootstrap approach that can be used to constrain the EST action beyond the Nambu-Got{\={o}} approximation~\cite{EliasMiro:2019kyf,Miro:2021rof}.
  
Indeed, it is by now clear that the Nambu-Got{\={o}} action should be considered only as a leading-order approximation of the actual EST describing the infrared dynamics of confining gauge theories. Going beyond this approximation is one of the most interesting open problems in this context. The terms beyond the Nambu-Got{\={o}} action encode important physical information and their study could be of great importance to understand the mechanism underlying confinement or the physical degrees of freedoms from which the confining string arises.

A natural way to study these corrections would be to perform high-precision simulations of the interquark potential in different lattice gauge theories. This approach, however, is hampered by the existence of boundary terms~\cite{Brandt:2010bw, Brandt:2017yzw, Brandt:2018fft, Brandt:2021kvt} related to the finiteness of the physical flux tubes studied in lattice simulations. These boundary terms provide a dominant contribution to the corrections beyond the Nambu-Got{\={o}} action and make the detection of other terms challenging (if possible at all: see below for a detailed discussion of this issue).

However, it can be shown that these corrections become subleading and can be neglected if one looks at the interquark potential at finite temperature, in the neighborhood of the deconfinement transition, but still in the confining phase. With this motivation, in this work we address a study of the simplest non-trivial non-Abelian lattice gauge theory: the $\SU(2)$ Yang-Mills theory in $(2+1)$ dimensions, in the range of temperatures $0.8 \Tc\leq T \leq \Tc$, where $\Tc$ denotes the deconfinement temperature.

This model has been the subject of several lattice studies in the past, including e.g. refs.~\cite{Ambjorn:1984me, Teper:1998te, Caselle:2004er, Caselle:2011vk, Bringoltz:2006zg, Brandt:2010bw, Brandt:2017yzw, Brandt:2018fft, Brandt:2021kvt, Bonati:2021vbc}, because it is a particularly simple lattice gauge theory based on a non-Abelian Lie group and allows one to study the non-perturbative features of Yang-Mills theories to much better numerical precision than one could obtain in lattice simulations of quantum chromodynamics.

There is another important reason why the choice of the $\SU(2)$ lattice gauge theory in $(2+1)$ dimensions is helpful. Since the deconfinement transition for this model is of the second order~\cite{Christensen:1990vc, Teper:1993gp}, renormalization group arguments show that in the neighborhood of the deconfinement transition the model is in the same universality class of the bidimensional Ising model: this is the celebrated Svetitsky-Yaffe conjecture~\cite{Svetitsky:1982gs}. According to this correspondence, the Polyakov loop correlator is mapped to the spin-spin correlator of the two-dimensional Ising model which, thanks to the exact integrability of the model, is exactly known. As will be shown in this work, the quantitative accuracy of this mapping is confirmed by our new sets of high-precision non-perturbative results, obtained by Monte~Carlo simulations; moreover, the knowledge of the exact form of the spin-spin correlator in the spin model provides us with a tool to extract the temperature-dependence of the ground-state energy of the theory with high accuracy, and to compare these results with effective-string predictions. In particular, this will allow us to precisely quantify the corrections with respect to the Nambu-Got{\={o}} action, which is one of the main goals of this work.

This article is organized as follows. The next four sections contain introductory material: section~\ref{sec:setup} is devoted to a description of the lattice setup, in section~\ref{sec:SY} we summarize the Svetitsky-Yaffe conjecture, section~\ref{sec:spin-spin_correlator_of_Ising_model} reviews known results on the spin-spin correlator of the two-dimensional Ising model, while section~\ref{sec:est} presents a brief introduction to the effective string theory. Our results and a description of lattice simulations are presented in section~\ref{sec:simulation_setting_and_results}, while in the last section~\ref{sec:concluding_remarks} we summarize our findings and list some concluding remarks.

\section{Definitions and lattice setup}
\label{sec:setup}

As mentioned above, in this work we focus on the three-dimensional $\SU(2)$ Yang-Mills theory at finite temperature. We regularize the theory on a finite cubic lattice of spacing $a$ and sizes $aN_t$ in the $\hat{0}$ (``Euclidean-time'') direction and $aN_s$ in the two other (``spatial'') directions, labelled as $\hat{1}$ and $\hat{2}$. To simplify notations, in the following we will set $a=1$. Periodic boundary conditions are assumed in all directions and we always take $N_s \gg N_t$. We use the standard Wilson action~\cite{Wilson:1974sk}
\begin{equation}
\SW = -\frac{2}{g^2} \sum_{x} \sum_{0 \le \mu < \nu \le 2} \Tr U_{\mu\nu} (x)
\end{equation}
where the plaquette $U_{\mu\nu}(x)=U_\mu(x)U_\nu\left(x+\hat{\mu}\right)U_\mu^\dagger\left(x+\hat{\nu}\right)U_\nu^\dagger(x)$ is defined as the path-ordered product of link variables $U_\mu(x)$ (taking values in the fundamental representation of the $\SU(2)$ group) along the elementary square having the site $x$ as a corner and lying in the oriented $(\mu,\nu)$ plane. $g^2$ is the squared bare coupling, which has energy dimension one. In the following, we will often use the parameter $\beta$, defined as $\beta=4/g^2$.

This lattice model has been studied in various works in the past. These include, in particular, ref.~\cite{Teper:1998te}, in which the following scale setting was presented:
\begin{equation}
\label{scale_setting}
\sqrt{\sigma_0(\beta)}=\frac{1.324(12)}{\beta}+\frac{1.20(11)}{\beta^2}+\mathcal{O}(\beta^{-3}) ,
\end{equation}
where $\sigma_0$ denotes the zero-temperature string tension.\footnote{Note that eq.~(\ref{scale_setting}) expresses the square root of the zero-temperature string tension as a dimensionless quantity, in lattice units. Reinstating the lattice spacing $a$, the left-hand side of eq.~(\ref{scale_setting}) would be replaced by $a\sqrt{\sigma_0}$} In the following, we will also discuss the relationship between $\sigma_0$ and the finite-temperature string tension $\sigma(T)$. For some of our simulations at $\beta=9.0$, we used the high-precision scale setting that was recently reported in ref.~\cite{Bonati:2021vbc}. \rev{In addition, we will also analyze the results of a set of simulations at $\beta=16.0$, that were presented in ref.~\cite{Athenodorou:2016kpd}.}

The temperature $T$ is related to the extent of the shortest compact size of the lattice as $N_t=1/T$: as a consequence, $T$ can be varied by changing $N_t$, or the lattice spacing (which can be varied continuously by tuning $\beta$), or both. We will study the system in the confining phase, just below the deconfinement temperature $\Tc$, in the temperature range $0.8\leq T/\Tc \leq 1$. Very accurate estimates of $\Tc$ for various values of $N_t$, that we will use in the following, can be found in ref.~\cite{Edwards:2009qw}.

The Polyakov loop through a point of spatial coordinates $\vec{x}$ is defined as the normalized trace of the closed Wilson line in the $\hat{0}$ direction:
\begin{equation}
P\left(\vec{x}\right) = \frac{1}{2} \Tr \prod_{t=0}^{N_t} U_0 \left(t,\vec{x}\right).
\end{equation}
The two-point correlation function of Polyakov loops is then defined as
\begin{equation}
\label{def_G}
G(R) = \left\langle \sum_{\vec{x}} \ P\left(\vec{x}\right) P\left(\vec{x}+R \hat{k}\right) \right\rangle ,
\end{equation}
where $\hat{k}$ denotes one of the two spatial directions, the sum is over all spatial coordinates $\vec{x}$, while the $\langle \dots \rangle$ average is taken over all values of all of the $U_\mu(x)$ variables, with a measure that is proportional to the product of the Haar measures of all $U_\mu(x)$ matrices and to $\exp(-\SW)$, and normalized in such a way that the expectation value of the identity operator is $1$.

\subsection{Finite-temperature interquark potential}

In a finite-temperature setting, one can define the ``interquark potential'' (or, more precisely, the potential energy associated with a pair of static fundamental color sources) $V$ from the free energy associated with the Polyakov-loop two-point correlation function:
\begin{equation}
G(R) \equiv \exp\left[-\frac{V(R,N_t)}{T}\right] = \exp\left[-N_t V(R,N_t)\right].
\label{polya}
\end{equation}
For sufficiently large values of the spatial separation $R$ between the color sources, in the confining phase one expects $V(R,N_t)$ to tend to a linearly rising function of $R$:
\begin{equation}
G(R) \simeq \exp\left[-\sigma(T) N_t R\right],
\label{area2}
\end{equation}
where $\sigma(T)$ denotes a temperature-dependent string tension. As we will see below, $\sigma(T)$ is a decreasing function of $T$ and vanishes exactly at the deconfinement point~\cite{Kaczmarek:1999mm,Cardoso:2011hh}. From $V(R,N_t)$ and $\sigma(T)$ it is also possible to derive the zero-temperature potential $V(R)$ as the $T\to 0$ limit of $V(R,N_t)$, and, accordingly, the zero-temperature string tension $\sigma_0$ as the $T\to 0$ limit of $\sigma(T)$.

It is interesting to note that the correlator defined in eq.~(\ref{polya}) has an analogy with the expectation value of an ordinary Wilson loop, except for the boundary conditions, which in this case are fixed in the spatial directions and periodic in the compact-time direction. Accordingly, the resulting geometry for the world-sheet associated with the fluctuating string between the color sources is that of a cylinder, which is topologically different from the rectangular geometry associated with a Wilson loop.

The existence of periodic boundary conditions for the gauge fields along the compactified Euclidean-time direction is the basis for the interpretation of the thermal deconfinement transition in terms of dynamical breaking of a global symmetry described by a group that is the center $C$ of the gauge group (i.e. $\Z_N$ if the gauge group is $\SU(N)$)~\cite{Polyakov:1978vu,Weiss:1980rj,McLerran:1981pb}. This symmetry, which can be thought of as the action of multiplying all timelike links at a given Euclidean time by the same element of $C$, has the Polyakov loop as an order parameter. In the low-temperature phase ($T<\Tc$), the center symmetry is realized, and the expectation value of a Polyakov loop vanishes: this means that the free energy associated with a static, isolated color source is finite, i.e. color confinement. Conversely, in the high-temperature phase ($T>\Tc$), center symmetry is spontaneously broken and the Polyakov loop has a non-zero expectation value: this signals that the free energy associated with a static color source is finite, i.e. that the theory is in its deconfined phase.

\section{Svetitsky-Yaffe mapping}
\label{sec:SY}

The peculiar role played by the Polyakov loops in the above discussion suggests to study the behavior of the theory in the vicinity of the deconfinement transition using an effective action that can be constructed by integrating out the spacelike link variables and projecting each Polyakov loop to the closest element of the center of the gauge group. Starting from a $(d+1)$-dimensional lattice gauge theory, we end up in this way with an effective action for the Polyakov loops, which will be a $d$-dimensional spin model, having the center of the original gauge group as a global symmetry.

This integration cannot be performed exactly, and one usually resorts to some kind of strong-coupling expansion. Notwithstanding this, some general insight on the behavior of the model can be deduced by simple renormalization group arguments~\cite{Svetitsky:1982gs}. Indeed, if the phase transition is continuous, in the vicinity of the critical point the fine details of the Hamiltonian describing the effective spin model can be neglected, and the latter can be shown to belong to the same universality class of the simplest spin model, with only nearest-neighbor interactions, sharing the same symmetry-breaking pattern. This means, in our case, that the deconfinement transition of the $\SU(2)$ lattice gauge theory in three dimensions, which is continuous, belongs to the same universality class of the symmetry-breaking phase transition of the two-dimensional Ising model. As is well known, this model is exactly integrable~\cite{Onsager:1943jn} and in particular, as we will see in the next section, an exact expression for the spin-spin correlator is known. This fact will play an important role in the following.

Let us list a few important features of the gauge-spin mapping suggested by the Svetitsky-Yaffe conjecture.
\begin{itemize}
\item The ordered (low-temperature) phase of the spin model corresponds to the deconfined (high-temperature) phase of the original gauge theory. These are the phases in which the order parameters of the two theories (namely, the Polyakov loop for the gauge theory and the spin for the spin model) have non-zero expectation values.  
\item The Polyakov loop is mapped to the spin operator, while the plaquette is mapped to the energy operator of the effective spin model. Accordingly, the Polyakov-loop correlator in the confining phase, in which we are interested in this work, is mapped to the spin-spin correlator in the disordered, high-temperature phase of the spin model.
\item Thermal perturbations from the critical point in the original gauge theory, which are driven by the plaquette operator, are mapped to thermal perturbations of the effective spin model, driven by the energy operator. Notice however the change in sign: an \emph{increase} in temperature of the original gauge theory corresponds to a \emph{decrease} of the temperature of the effective spin model.
\end{itemize}
An important consequence of this mapping is that, in the vicinity of the deconfinement point, we can use the spin-spin correlator of the Ising model to model the behavior of the Polyakov loop correlator of the lattice theory: this poses very tight constraints on the EST that describes this correlator. In this respect, the exact integrability of the two-dimensional Ising model gives us a unique opportunity to study the dynamics of the $\SU(2)$ lattice gauge theory in $(2+1)$ dimensions.

\section{The spin-spin correlator of the two-dimensional Ising model}
\label{sec:spin-spin_correlator_of_Ising_model}

The spin-spin correlator of the two-dimensional Ising model can be written as a finite determinant with a size that depends on the separation of the spins~\cite{McCoy_book}. In the scaling limit, these determinants can be rewritten as suitable solutions of an equation of the Painlev\'e type~\cite{Wu:1975mw}. These solutions have a very different form depending on the phase of the model. In particular, denoting the correlation length by $\xi$ and the $R/\xi$ ratio as $t$, in the disordered phase which is the focus of our interest in this work they can be expanded in the short-distance ($R\ll\xi$) and in the long-distance ($R\gg\xi$) limits as follows.
\begin{itemize}
\item In the $R\ll\xi$ regime, the two-point spin correlator can be expanded as
\begin{equation}
\label{shortR}
    \left\langle{\sigma(0)\sigma(R)}\right\rangle = \frac{k_s}{R^\frac{1}{4}}\bigg[1+\frac{t}{2} \ln\bigg(\frac{e^\gammaE t}{8}\bigg)+\frac{1}{16}t^2+\frac{1}{32}t^3\ln\bigg(\frac{e^\gammaE t}{8}\bigg)+O(t^4\ln^2t)\bigg],
\end{equation}
where $\gammaE=0.57721\dots$ denotes the Euler-Mascheroni constant, while $k_s$ is a non-universal constant, which can be evaluated exactly in the case of the two-dimensional Ising model on a square lattice~\cite{Wu:1975mw} (a results which, due to its non-universal nature, is not of relevance for our present purposes).
\item Conversely, in the $R\gg\xi$ regime, the two-point spin correlator admits the expansion
\begin{equation}
\label{largeR}
    \left\langle{\sigma(0)\sigma(R)}\right\rangle = {k_l} K_0(t)
\end{equation}
where, again, $k_l$ is a non-universal constant which can be evaluated exactly in the case of the two-dimensional Ising model on a square lattice (but which is not relevant for our discussion), while $K_0$ is the modified Bessel function of order zero, whose long-distance expansion is
\begin{equation}
\label{K0_expansion}
K_0(t)\simeq \sqrt{\frac{\pi}{2t}}e^{-t}\left[ 1 + O\left(\frac{1}{t}\right)\right].
\end{equation}
\end{itemize}
We conclude this section with some important observations.
\begin{enumerate}
\item The short-distance expansion of eq.~(\ref{shortR}) can also be obtained using conformal perturbation theory (CPT): an approach proposed in 1987 by Zamolodchikov~\cite{Zamolodchikov:1987ti}, which for the past thirty years has proved to be a powerful analytical tool to describe statistical-mechanics models and quantum field theories in the vicinity of a critical point. As a matter of fact, the full agreement of the CPT result in the Ising case with the exact expansion of eq.~(\ref{shortR}) represents one of the most stringent and successful tests of CPT~\cite{Mikhak:1993py,Mikhak:1993vf, Guida:1995kc}. While original applications of CPT were limited to two-dimensional models (see, for example, refs.~\cite{Zamolodchikov:1990bk, Guida:1995kc, Guida:1996ux, Caselle:1999mg, Amoretti:2017aze, Amoretti:2020bad}), recently it has also been extended to three-dimensional models~\cite{Caselle:2015csa, Caselle:2016mww}. The reason why the CPT approach is important for our present discussion is that, thanks to it, the results that we discuss in this article do not necessarily require the exact integrability of the underlying spin model but can be extended to any pair of lattice gauge theory and spin model with a second-order deconfinement/symmetry-breaking transition. Indeed this approach was already followed in ref.~\cite{Caselle:2019tiv} for the mapping between the $(3+1)$-dimensional $\SU(2)$ Yang-Mills theory and the Ising model in three dimensions.
\item The $K_0$ function describing the long-distance behavior is the typical expression that one obtains for the correlator of a two-dimensional Euclidean quantum field theory with an isolated, massive excitation. In the Ising case, this is the Majorana fermion which describes the model in the continuum limit. As for the short-distance expansion, also this result can be extended to models that are not exactly integrable. Indeed the symmetric phase of a generic spin model is always described, in the continuum limit, by an appropriate set of (possibly interacting) massive particles. In the long-distance limit, the spin-spin correlator for any of these theories will be dominated by the lowest mass in the spectrum,  whose correlation function in $d$ dimensions is given by a $K_{(d-2)/2}$ modified Bessel function.
\item It is interesting to observe the shift in the exponent of the $1/R$ term in eq.~(\ref{shortR}) and in eq.~(\ref{largeR}). At short distance (where we expect deviations with respect to the Nambu-Got{\={o}} action) the power is fixed by the universality class of the model (in our case $1/4$), and, in general, will depend on the gauge group of the underlying gauge theory. At large distances, instead, the asymptotic behavior of the Bessel function in eq.~(\ref{K0_expansion}) implies that the power is always $1/2$: this is in agreement with the Nambu-Got{\={o}} result, as we will discuss in section~\ref{sec:est}.
\end{enumerate}

\section{Effective string theory predictions}
\label{sec:est}

There is by now a rich literature on the many properties and features of the effective string theory description of the confining flux tube. Here we will limit ourselves to a brief discussion of the features that are most relevant for our problem. For a general introduction to EST, we refer the reader to the recent reviews~\cite{Aharony:2013ipa,Brandt:2016xsp,Caselle:2021eir}.

The main idea behind EST is that confinement of color charges can be associated with the formation of a thin string-like flux tube~\cite{Nambu:1974zg,Goto:1971ce,Luscher:1980ac,Luscher:1980fr,Polchinski:1991ax}, which leads, for large separations between the color sources, to a linearly confining potential.

In a finite-temperature setting it is possible to show, with very mild assumptions, that the EST description implies the following form for the Polyakov-loop correlator~\cite{Meyer:2006qx, Luscher:2004ib}:
\begin{equation}
\label{EST}
    \left\langle{P(0)P^\dagger(R)}\right\rangle = \sum_n |v_n(N_t)|^2 2R\bigg(\frac{E_n}{2\pi R}\bigg)^\frac{D-1}{2}K_{(D-3)/2}(E_nR),
\end{equation}
where $D$ denotes the number of spacetime dimensions (in our case $D=3$), while $E_n$ are the energy levels of the string and $v_n(N_t)$ their amplitudes, which in general depend on the inverse temperature $N_t$. The physical meaning of eq.~(\ref{EST}) is that the Polyakov-loop correlation function can be modelled in terms of an infinite series of modified Bessel functions of the second kind: this is expected to hold independently from the type of string that is considered, as long as the energy spectrum is characterized by the existence of isolated states, and these states are stable against decay by glueball radiation. 

At large $R$, the right-hand side of eq.~(\ref{EST}) is dominated by the lowest energy level $E_0$ and, setting $D=3$, we end up with exactly the same expression that we found for the long-distance behavior of the spin-spin correlator in the two-dimensional Ising model, eq.~(\ref{largeR}). It is interesting to note that this equality is not simply a consequence of the $\SU(2)$/Ising correspondence that we are studying here: instead, it is much more general and holds also for $D>3$~\cite{Caselle:2021eir}.  As we mentioned above, any spin model with an isolated ground state in the spectrum is described by a modified Bessel function of the same type that appears in the EST description. This is an important consistency check of both the EST picture\footnote{In particular, as we will see below, for the Nambu-Got{\={o}} action, both $v(N_t)$ and $E_n$ can be evaluated exactly thanks to the exact integrability of the model.} and of the Svetitsky-Yaffe mapping in this high temperature limit.

The Nambu-Got{\={o}} string model~\cite{Nambu:1974zg,Goto:1971ce} is the simplest Poincar\'e invariant EST. It has a simple geometric interpretation, since it associates each possible configuration that the string can span in the target space with a quantum weight proportional to the area of the world-sheet surface. As such, the Nambu-Got{\={o}} string action can be thought of as a straightforward generalization of the relativistic action for a pointlike particle to a bosonic string. The Nambu-Got{\={o}} action can be written as follows:
\begin{align}
\label{NGaction}
S_{\mbox{\tiny{NG}}}= \sigma_0 \int_\Sigma d^2\xi \sqrt{g},
\end{align} 
where $g\equiv \det g_{\alpha\beta}$ and $g_{\alpha\beta}=\partial_\alpha X_\mu~\partial_\beta X^\mu$ is the induced metric on the reference world-sheet surface $\Sigma$, where we denote the world-sheet coordinates as $\xi\equiv(\xi^0,\xi^1)$. This action has only one free parameter: the string tension $\sigma_0$, which has dimension two. 

In order to perform calculations with the Nambu-Got{\={o}} action one has first to fix its invariance under reparametrizations of the string world-sheet coordinates. A common choice is the so-called ``physical gauge'', in which the two world-sheet coordinates are identified with the longitudinal degrees of freedom of the string: $\xi^0=X^0$ and $\xi^1=X^1$, so that the string action can be expressed as a function only of the $(D-2)$ degrees of freedom describing transverse displacements, $X^i$, with $i=2, \dots , (D-1)$, which are assumed to be single-valued functions of the world-sheet coordinates. In the physical gauge, the determinant of the metric has the form
\begin{equation}
g=1+\partial_0 X_i\partial_0 X^i+\partial_1 X_i\partial_1 X^i+\partial_0 X_i\partial_0 X^i\partial_1 X_j\partial_1 X^j-(\partial_0 X_i\partial_1 X^i)^2
\end{equation}
and the Nambu-Got{\={o}} action can then be written as a low-energy expansion in the number of derivatives of the transverse degrees of freedom of the string which, by a suitable redefinition of the fields, can be rephrased as an expansion around the limit of an infinitely long string. The first few terms in this expansion are 
 \begin{equation}
S=\sigma_0RN_t+\frac{\sigma_0}{2}\int d^2\xi\left[\partial_\alpha X_i\cdot\partial^\alpha X^i+
\frac18(\partial_\alpha X_i \cdot\partial^\alpha X^i)^2
-\frac14(\partial_\alpha X_i \cdot\partial_\beta X^i)^2+\dots\right].
\label{action2NG}
\end{equation}

\rev{It is important to stress, however, that the physical gauge discussed above is anomalous in $D\not= 26$. Hence, in the three-dimensional case we are interested in, eq.~(\ref{action2NG}) only describes an {\sl effective} version of the original Nambu-Got{\={o}} action. However, thanks to the low-energy universality theorem discussed below, it is known that the corrections to eq.~(\ref{action2NG}) due to the anomaly only appear at high orders in the low-energy expansion. Finding the leading corrections with respect to the physical-gauge limit is one of the goals of this paper.} 

Despite its apparent complexity, it can be shown that all the additional terms in the expansion of eq.~(\ref{action2NG}) beyond the Gau{\ss}ian one conspire to yield an exactly integrable, irrelevant perturbation of the Gau{\ss}ian term~\cite{Dubovsky:2012sh}, driven by the $T\overline{T}$ operator of $D-2$ free bosons~\cite{Caselle:2013dra}.

Thanks to this exact integrability, the partition function of the model can be written explicitly.\footnote{\rev{The explicit expression for the partition function was actually found even before this $T\overline{T}$ study,  first by using the constraints imposed by the open-closed string duality~\cite{Luscher:2004ib} and then using a D$p$-brane formalism~\cite{Billo:2005iv}. In the context of the $T\bar T$ formalism it was recently described in refs.~\cite{Aharony:2018bad,Datta:2018thy} for periodic boundary conditions, and in ref.~\cite{Beratto:2019bap}, for the Dirichlet boundary conditions of relevance for our study. In addition, the latter reference also studied the extension to compactified transverse dimensions.}} For the two-point Polyakov-loop correlation function that we are considering here \rev{(i.e. for the partition function with Dirichlet boundary conditions in the $R$ direction and periodic boundary conditions in the $N_t$ direction)}, the expression in $D$ spacetime dimensions is
\begin{equation}
  G(R) =\sum_{n=0}^{\infty}w_n\frac{2R \sigma_0 N_t}{{E}_n}
   \left(\frac{\pi}{\sigma_0}\right)^{\frac{D-2}{2}}
  \left(\frac{{E}_n}{2\pi R}\right)^{\frac{D-1}{2}}
  K_{(D-3)/2}({E}_nR),
\label{NGPP}
\end{equation}
where the energy levels $E_n$ are given by
\begin{equation}
  {E}_n=\sigma_0 N_t
  \sqrt{1+\frac{8\pi}{\sigma_0 N_t^2}\left(n-\frac{D-2}{24}\right)}.
\label{energylevels}
\end{equation}
and the weights $w_n$ can be obtained from the expansion in powers of $q$ of the Dedekind function that describes the large-$R$ limit of eq.~(\ref{NGPP}) (for a detailed derivation, see ref.~\cite{Billo:2005iv}):
\begin{equation}
\label{etaexp}
\left(\prod_{r=1}^\infty \frac{1}{1 - q^r}\right)^{D-2}
= \sum_{k=0}^\infty w_k q^k.
\end{equation}
For $D=3$ we have simply $w_k=p_k$, the number of partitions of the integer $k$. Similar expressions can be obtained also for the other geometries, e.g. for the rectangle (relevant for the description of the Wilson loop)~\cite{Billo:2011fd} and for the torus (which can model an interface)~\cite{Billo:2006zg}.

From the discussion above we see that, as anticipated, \rev{for the effective Nambu-Got{\={o}} string} we have an exact expression both for the $v(N_t)$ amplitudes\rev{\footnote{We remark that this expression for the amplitudes is correct only in the effective-string limit (i.e. assuming the physical gauge), as also the amplitudes are affected by the anomaly. As we only study the ground-state energy $E_0$, we do not consider this issue further in the following.}} and for the energy levels $E_n$. In particular, for the $D=3$ case the lowest state is 
\begin{equation}
E_0=\sigma_0 N_t
  \sqrt{1-\frac{\pi}{3\sigma_0 N_t^2}} = \sigma(T) N_t ,
\label{E0}
\end{equation}
where we defined the ``temperature-dependent string tension'' as
\begin{equation}
\sigma(T)  \equiv \sigma_0 \sqrt{1-\frac{\pi}{3\sigma_0 N_t^2}} .
\label{sigmaT}
\end{equation} 
As we have seen in the previous section, $E_0$ is the inverse of the correlation length, thus the Nambu-Got{\={o}} EST predicts a critical temperature~\cite{Olesen:1985ej,Pisarski:1982cn}
\begin{equation}
\frac{T_{c,NG}}{\sqrt{\sigma_0}}=\sqrt{\frac{3}{\pi(D-2)}}
\end{equation}
and a critical index $\nu=1/2$. This prediction, however, is inconsistent with the Svetitsky-Yaffe mapping, which for our model predicts the two-dimensional Ising value $\nu=1$. Moreover, the prediction for the critical temperature is quantitatively wrong (albeit close to the correct one). These observations suggest that, in order to obtain the correct EST describing the gauge theory, one should necessary go beyond the Nambu-Got{\={o}} approximation.

\subsection{Effective string action beyond the Nambu-Got{\={o}} approximation}
\label{beyond}

The discussion above shows that the pure Nambu-Got{\={o}} action cannot be the actual effective string action. Discovering the correct (subleading) terms of the effective string action beyond the Nambu-Got{\={o}} approximation is one of the major open challenges in present studies of the EST, and is, in fact, the main goal of this article.
 
There are essentially two classes of terms which one should address: ``bulk terms'' and ``boundary terms''.  Let us look at them in more detail.

\subsubsection{Beyond the Nambu-Got{\={o}} approximation: bulk terms and low-energy universality}

From an effective-action point of view, there is no reason to constrain the coefficients of the higher-order terms in eq.~(\ref{action2NG}) to the values they take in the derivative expansion of the Nambu-Got{\={o}} action. One should instead assume the most general form for such an effective action:
\begin{equation}
S=S_{\mbox{\tiny{cl}}}+\frac{\sigma_0}{2}\int d^2\xi\left[\partial_\alpha X_i\cdot\partial^\alpha X^i+
c_2(\partial_\alpha X_i \cdot\partial^\alpha X^i)^2
+c_3(\partial_\alpha X_i \cdot\partial_\beta X^i)^2+\dots\right],
\label{action2}
\end{equation}
and then fix the coefficients (which, in this context, would play the role of low-energy constants of the effective theory) order by order, either using results from Monte~Carlo simulations or (in the case of quantum chromodynamics) from experiments.

However, one of the most interesting results of the last few years, known as ``low-energy universality of the EST'' is that the $c_i$ coefficients are not arbitrary, but must satisfy a set of constraints to enforce the Poincar\'e invariance of the gauge theory in the target space~\cite{Meyer:2006qx,Luscher:2004ib, Aharony:2009gg, Aharony:2011gb, Gliozzi:2011hj, Gliozzi:2012cx}. This same result can also be obtained in an independent way, using a bootstrap type of analysis: this was done in refs.~\cite{EliasMiro:2019kyf,Miro:2021rof} in the framework of the S-matrix approach pioneered in ref.~\cite{Dubovsky:2012sh}. These constraints are particularly restrictive for a three-dimensional theory: the first few terms of the expansion exactly coincide with those that are obtained from the expansion of the Nambu-Got{\={o}} action, while the first correction with respect to the Nambu-Got{\={o}} action, in the high-temperature regime which we are studying here, appears only at order $1/N_t^7$. Moreover, even the coefficient of this correction can be constrained: using the notations of refs.~\cite{EliasMiro:2019kyf,Miro:2021rof}, this additional term can be written as
\begin{equation}
-\frac{32\pi^6}{225}\frac{\gamma_3}{\sigma^3N_t^7}
\label{gamma3}
\end{equation}
where $\gamma_3$ is a new parameter which together with $\sigma_0$ defines the EST. By using 
a bootstrap analysis it is possible to show that $\gamma_3$ is constrained to be $\gamma_3>-\frac{1}{768}$. 

The $\gamma_3$ parameter encodes some important information on the effective string theory. For example, it can be shown that \rev{only if $\gamma_3\geq 0$ then the Axionic String \emph{Ansatz} (ASA) discussed in refs.~\cite{Dubovsky:2013gi,Dubovsky:2014fma} is certainly correct, while nothing can be said on it if $\gamma_3<0$}.

\subsubsection{Beyond the Nambu-Got{\={o}} approximation: boundary corrections}
\label{subsubsect:boundary}

Boundary corrections to the EST encode the effect of possible interactions of the flux tube with the static color sources at its ends. As we will see below, at zero and very low temperature the boundary correction behaves as $1/R^4$, and hence, is the dominant contribution beyond the Nambu-Got{\={o}} approximation. Its presence makes it almost impossible to detect the much weaker (and more interesting, being related to the nature of the confining flux tubes) effects due to bulk correction terms discussed above. However, as will be shown below, in the high-temperature regime which we studied in this work, the boundary term actually becomes \emph{subleading}, making it possible to access the bulk corrections using Monte~Carlo simulations.

Like the bulk terms, also the boundary terms are strongly constrained by Lorentz invariance: the first boundary correction compatible with the spacetime symmetries of the underlying gauge theory is~\cite{Billo:2012da}
\begin{equation}
b_2\int d\xi_0 \left[
\frac{\partial_0\partial_1 X\cdot\partial_0\partial_1 X}{1+\partial_1 X\cdot\partial_1X}-
\frac{\left(\partial_0\partial_1 X\cdot\partial_1 X\right)^2}
{\left(1+\partial_1 X\cdot\partial_1X\right)^2}\right],
\label{firstb}
\end{equation} 
with an arbitrary, non-universal coefficient $b_2$. This coefficient has been estimated in some recent lattice studies~\cite{Brandt:2010bw,Brandt:2017yzw,Brandt:2018fft,Brandt:2021kvt,Billo:2012da}: in particular, for the $\SU(2)$ Yang-Mills theory in three dimensions it was found to be $b_2\simeq-0.025/(\sqrt{\sigma})^3$. The lowest-order term of the expansion of eq.~(\ref{firstb}) is
\begin{equation}
\label{derexpsb1}
S_{b,2}^{(1)} = b_2\int d\xi_0 (\partial_0\partial_1 X)^2.
\end{equation}
The contribution of this term to the interquark potential was evaluated in ref.~\cite{Aharony:2010cx} using a $\zeta$-function regularization:
\begin{equation}
\label{polybound}
\langle S^{(1)}_{b,2} \rangle=-b_2\frac{\pi^3 N_t}{60 R^4} E_4(e^{-\frac{\pi N_t}{ R}})
\end{equation}
where $E_4$ denotes the fourth-order Eisenstein series:
\begin{equation}
E_{4}(q)\equiv  1+\frac{2}{\zeta(-3)}\sum_{n=1}^{\infty} \frac{n^{3}q^n}{1-q^n}~
\end{equation}
and $\zeta(s)$ is the Riemann $\zeta$ function.

In the low-temperature ($N_t\gg R$) regime (which is the one that is most often studied in lattice calculations) eq.~(\ref{polybound}) amounts to a $1/R^4$ contribution to the interquark potential. As it scales with a larger power of $R$, this term obfuscates the evidence of bulk corrections in numerical results. However, using the modular properties of the Eisenstein function,
\begin{equation}
E_4\left(e^{-\frac{\pi N_t}{R}}\right)=\left(\frac{2R}{N_t}\right)^4 E_4\left(e^{-\frac{4 \pi R}{N_t}}\right),
\end{equation}
it is easy to see that in the $R\gg N_t$ (``high-temperature'') regime, the boundary correction becomes 
\begin{equation}
\label{polybound2}
\langle S^{(1)}_{b,2} \rangle=-b_2\frac{4 \pi^3}{15 N_t^3} E_4\left(e^{-\frac{4\pi R}{N_t}}\right),
\end{equation}
which does not contain terms linear in $R$ and thus it does not contribute to the temperature-dependent string tension. We will make use of this property in the analysis of our numerical results.

\section{Simulation setting and results}
\label{sec:simulation_setting_and_results}

In this section we present the results of a new set of Monte~Carlo simulations. The calculations were run with the parallel C++ code developed for the studies presented in refs.~\cite{Panero:2009tv,Mykkanen:2012ri}. The elements of the $\SU(2)$ group are stored as four complex numbers in double precision, and are updated using a combination of local heat-bath~\cite{Creutz:1980zw, Kennedy:1985nu} and overrelaxation~\cite{Adler:1981sn} steps.

We performed two sets of simulations. In the first, which was mainly devoted to testing the Svetitsky-Yaffe mapping and to a general study of the deviations with respect to the Nambu-Got{\={o}} EST predictions, we fixed a few values of $N_t$ and varied the temperature by changing $\beta$. In the second set of simulations, which was aimed to a high-precision study of the corrections to the Nambu-Got{\={o}} action, we chose the opposite strategy and fixed three values of $\beta$ and varied the temperature by changing $N_t$. Let us discuss these simulations in detail.

\subsection{Test of the Svetitsky-Yaffe mapping}

We chose four values of $N_t$, namely $N_t=6$, $7$, $8$, and $9$, for which a very precise determination of $\betac$ is known from ref.~\cite{Edwards:2009qw}, and we performed a large set of simulations for different values of $\beta$,
see tab.~\ref{table_1}. \rev{We chose the values of $\beta$ so as to keep the correlation length $\xi$ in the range $10 \lesssim \xi \lesssim 25$ lattice spacings, in order to control both lattice artifacts and finite-size effects.}

\begin{table}[h]
\centering
    \begin{tabular}{ cc }
    \begin{tabular}{ |c|c|c|c| } 
    \hline
    $N_t\times N_s^2$ & $\beta$ & $T/\Tc$ & $\nconf$ \\ 
    \hline \hline
    \multirow{2.8}{*}{$9\times96^2$} & $11.3048$ & $0.80$ & $2.5\times10^5$ \\
    \cline{2-4}
    & $11.72873$ & $0.83$ & $2.5\times10^5$ \\
    \cline{1-4}
    $9\times160^2$ & $12.15266$ & $0.86$ & $2.5\times10^5$ \\
    \hline 
    \multirow{2.8}{*}{$7\times96^2$} & $9.228023$ & $0.83$ & $2.5\times10^5$ \\
    \cline{2-4}
    & $9.561566$ & $0.86$ & $2.5\times10^5$ \\
    \hline
    \end{tabular} &
    \begin{tabular}{ |c|c|c|c| } 
    \hline
    $N_t\times N_s^2$ & $\beta$ & $T/\Tc$ & $\nconf$ \\ 
    \hline \hline
    \multirow{4.3}{*}{$8\times96^2$} & $10.10736$ & $0.80$ & $2.5\times10^5$ \\
    \cline{2-4}
    & $10.486386$ & $0.83$ & $2.5\times10^5$ \\
    \cline{2-4}
    & $10.865412$ & $0.86$ & $2.5\times10^5$ \\
    \hline 
    \multirow{2.8}{*}{$6\times96^2$} & $8.258494$ & $0.86$ & $2.5\times10^5$ \\
    \cline{2-4}
    & $8.546581$ & $0.89$ & $2.5\times10^5$ \\
    \hline
    \end{tabular}
    \end{tabular}
    \caption{Information on the first set of lattice simulations.}
    \label{table_1}
\end{table}

\begin{figure}[!htb]
\centering
\includegraphics[width=0.95\textwidth, clip]{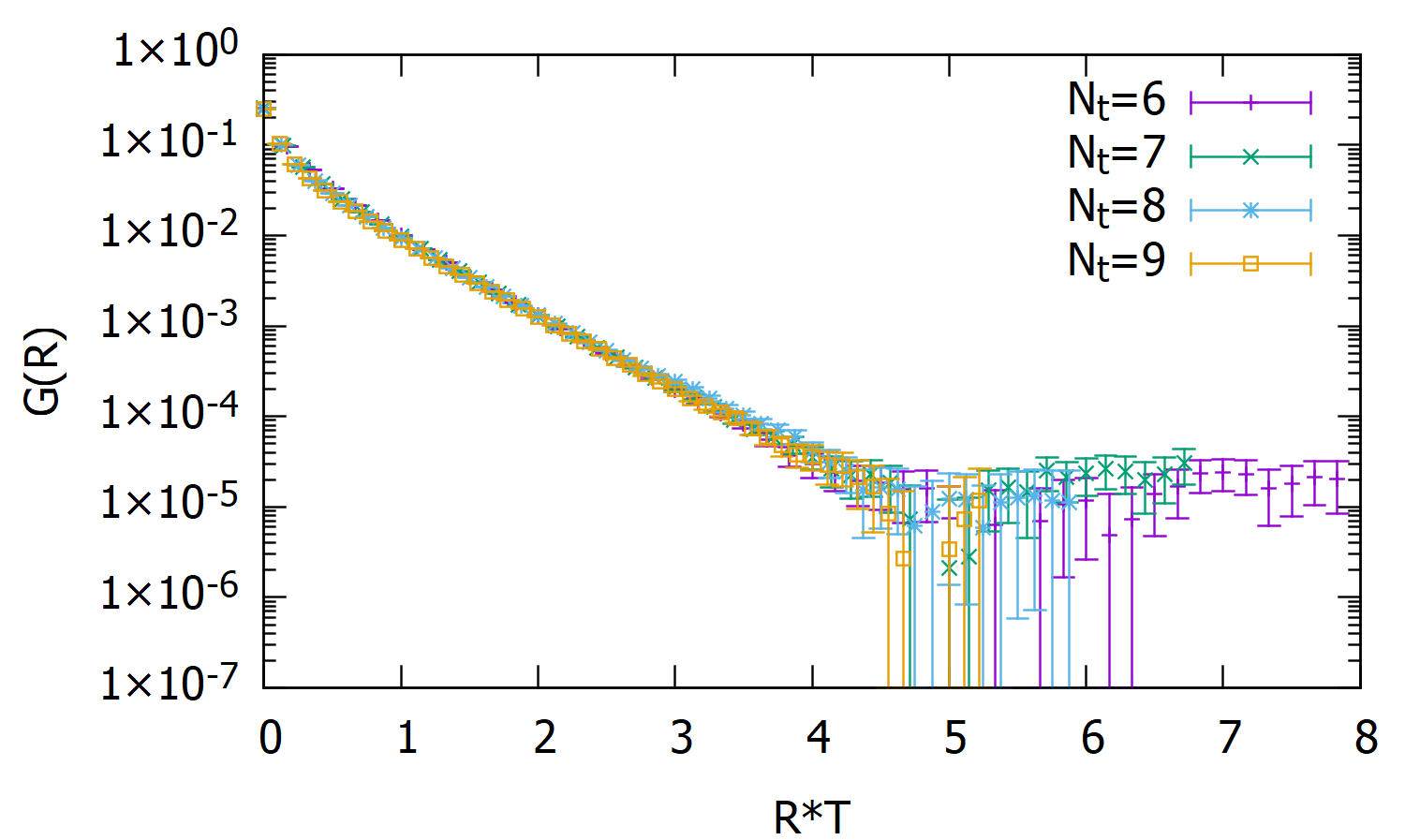}
\caption{Collapse of data in the Polyakov-loop two-point correlation function $G(R)$ obtained for different values of $N_t$, from $6$ to $9$, at the same temperature, $T=0.62 \Tc$. The data, shown using a logarithmic scale for the vertical axis, are plotted as a function of the distance between the loops, $R$, in units of the inverse temperature.}\label{fig_1}
\end{figure}

In addition, in fig.~\ref{fig_1} we present a data sample corresponding to $T=0.62 \Tc$ \rev{(a temperature lower than the rest of those considered in this work, that corresponds to a shorter correlation length and allows a better visualization of data collapse),} again from simulations with $N_t=6$, $7$, $8$, and $9$: the figure, showing the Polyakov loop correlator as a function of the distance (in units of the inverse temperature) reveals a clear collapse of data. The decay of the correlator can be described very well in terms of a single exponential (which, in this semilogarithmic plot, manifests itself in the approximately linear behavior of the data) over a wide range of distances. Leaving aside the points at values of $R$ of the order of a few lattice spacings, which are affected by non-negligible discretization effects, the slight bending of the data from short to intermediate distances, before the onset of the purely exponential decay, is a signature of the effective string corrections that will be discussed in detail below.
 
\rev{In the analysis of these correlators, one must pay particular attention to finite-size effects, which could become particularly severe when the correlation length grows as the critical temperature is approached. Fortunately, the knowledge of the exact solution of the Ising model in two dimensions allows us to test the Svetitsky-Yaffe mapping in the short-distance regime, using eq.~(\ref{shortR}), which is less sensitive to finite-size corrections. For the subset of our data sets corresponding to correlation lengths shorter than $N_s/6$, we found that finite-size effects could be accounted for by including the contribution from the first periodic copy of the system in the fit, enabling us to cross-check our results using \emph{also} the long-distance behavior of the Ising correlator. The role of finite-size corrections will be analyzed more in detail below, in the discussion of our determination of the $\gamma_3$ coefficient.}

In the following, we will discuss in detail only one of our simulations, the one at $N_t=9$ and $\beta=12.15266$, \rev{corresponding to $T/\Tc=0.86$ and $\xi\simeq 22$, for which we could test the Svetitsky-Yaffe mapping both at short and at long distances. To control finite-size effects we simulated the model on a lattice with $N_s=160$. We found analogous results for all the other cases when we could perform both fits, as reported in tab.~\ref{table_confront}.}

\begin{figure}[!htb]
  \centering
  \includegraphics[width=0.95\textwidth, clip]{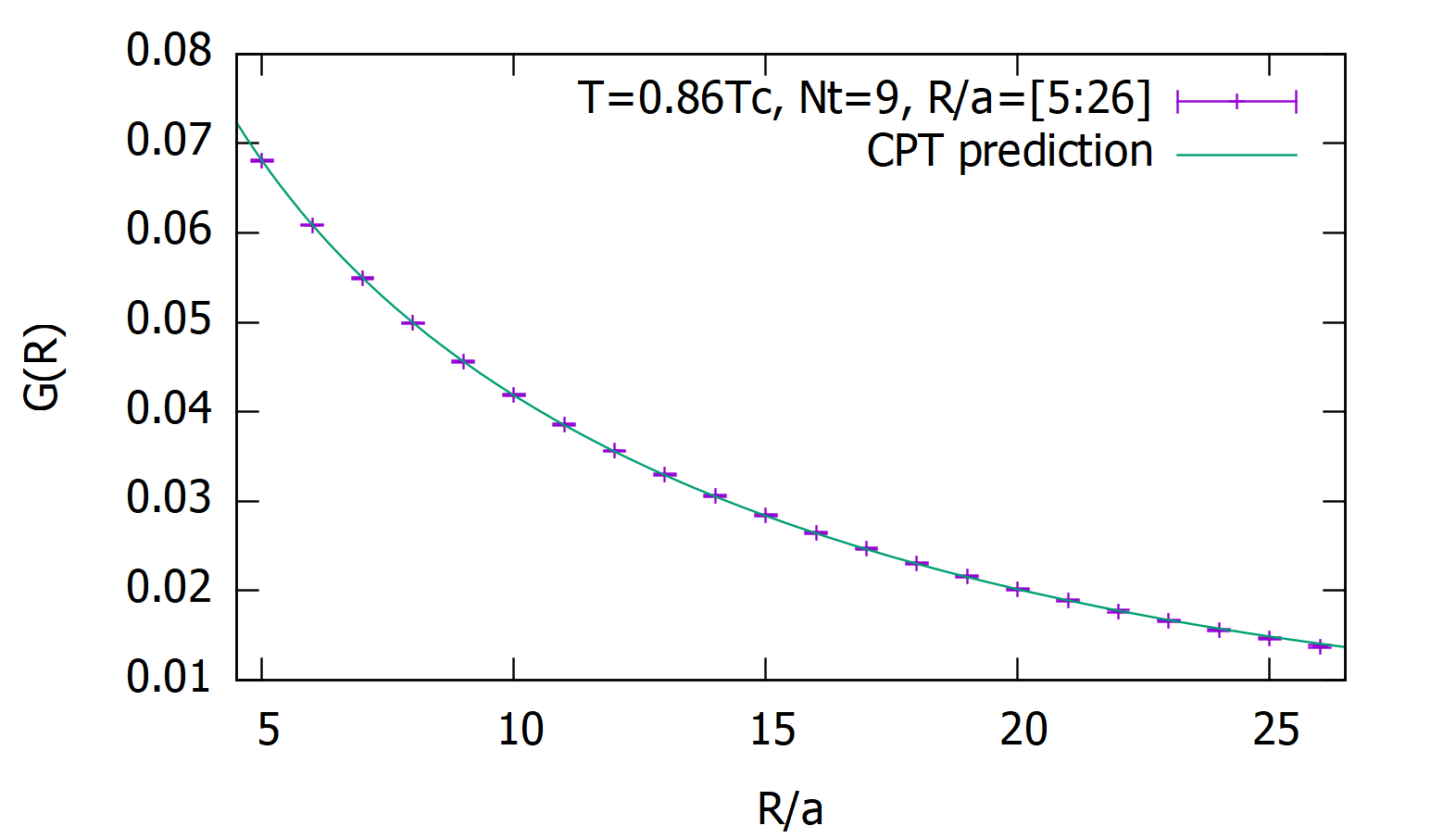}
  \caption{Fit of the Monte~Carlo results obtained at $N_t=9$, $\beta=12.15266$, $T/\Tc=0.86$ according to the short-distance approximation of the spin-spin Ising correlator of eq.~(\ref{shortR}).}\label{fig_2}
\end{figure}

\begin{figure}[!htb]
  \centering
  \includegraphics[width=0.95\textwidth, clip]{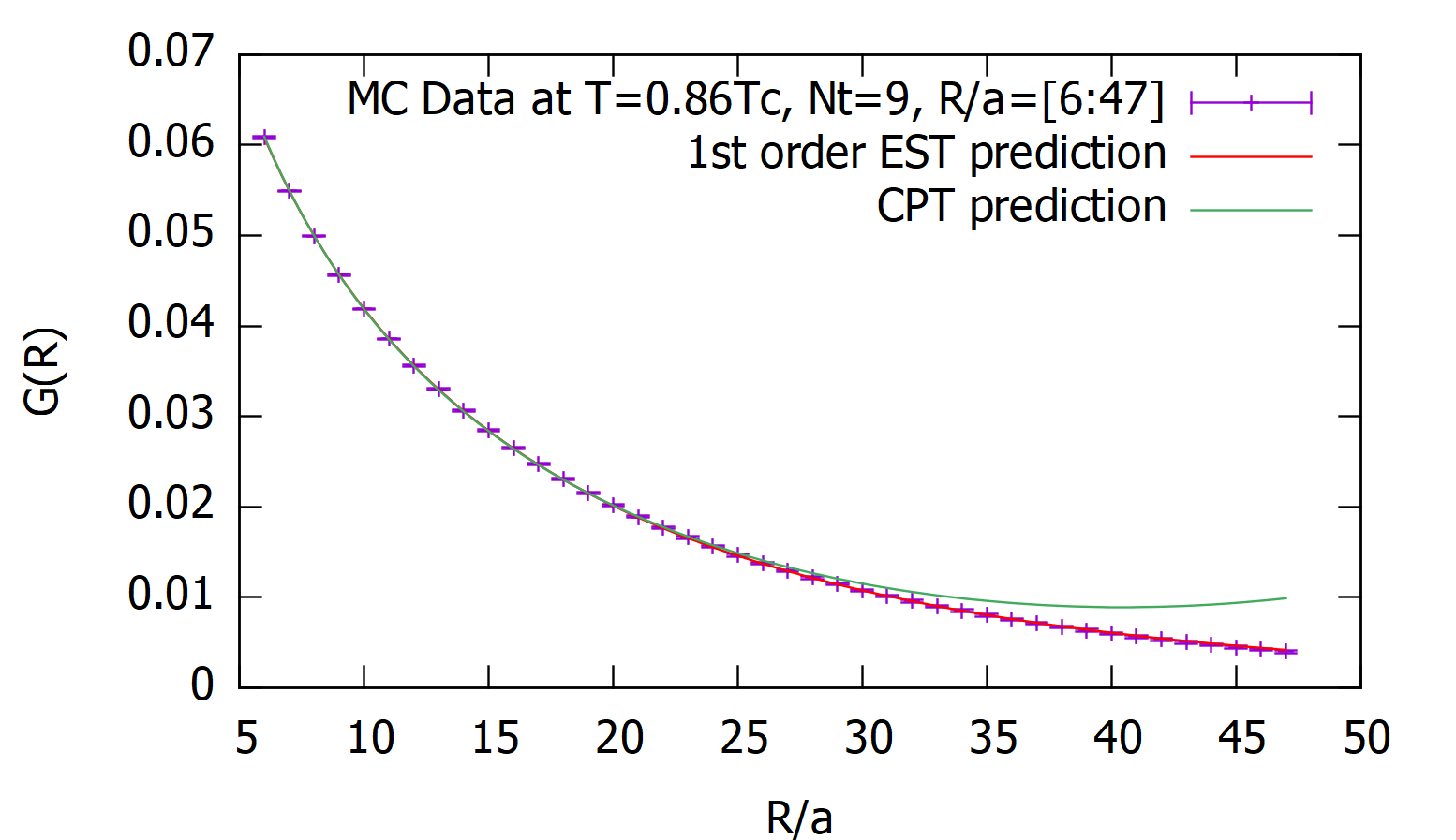}
  \caption{Fit of the data at $N_t=9$, $\beta=12.15266$, $T/\Tc=0.86$ combining both the short- and long-distance approximations of the spin-spin Ising correlator, which are respectively given by eq.~(\ref{shortR}) and by eq.~(\ref{largeR2}).}\label{fig_3}
\end{figure}

We first tested the Svetitsky-Yaffe mapping, fitting our data to the Ising expressions for the spin-spin correlator reported in eq.~(\ref{shortR}) and in eq.~(\ref{largeR}). To account for the first periodic copy of the lattice, the long-distance fits were performed with the following function 
\begin{equation}
G(R) = {k_l} \left[K_0\left(\frac{R}{\xi}\right)+K_0\left(\frac{N_s-R}{\xi}\right)\right]
\label{largeR2}
\end{equation}
which, like eq.~(\ref{largeR}), has only two free parameters: $k_l$ and $\xi$. The results of the fits are listed in tab.~\ref{table:2} and shown in fig.~\ref{fig_2} and in fig.~\ref{fig_3}.  
 
From tab.~\ref{table:2} it is clear that both at short and at long distances we could fit a wide range of data with a good reduced $\chi^2$.
\rev{As expected, the short-distance fit inspired by the Ising-model correlation function works well only up to distances of the order of the correlation length ($R\in [5-26]$).
What is instead surprising is that the long-distance fit works all the way down to $R=6$, exhibiting remarkable agreement with the simulation results. This agreement holds also for the other values of $\beta$ that we tested.  What is more important, 
the two estimates of the correlation length agree with each other within their uncertainties. This means that the two-point Polyakov-loop correlator is described well by the Ising spin-spin correlation function in the whole range of distances that we studied (down to $R>4$). 
It is tempting to guess that the good quality of the long-distance fit is due to the fact that, as mentioned above, the fitted function coincides with the EST prediction if one neglects all higher-order states of the spectrum except for $E_0$. In this respect it is worth noting that the minimum value of $R$ that can be described by the fit almost coincides with the critical radius $R_c$ at which the Nambu-Got{\={o}} action is expected to break down due to the tachyonic singularity~\cite{Aharony:2013ipa}, which at $\beta=12.15266$ is $R_c=\sqrt{\frac{\pi}{12\sigma}}\simeq 4.4$.
Since higher-order states in the EST expression cannot be detected within the precision of our simulations, we cannot use them to extract information on the corrections beyond Nambu-Got{\={o}}. To obtain information on these corrections, it is more convenient to use a different strategy, that will be discussed in the next subsection.

}

\begin{table}[h]
\centering
    \begin{tabular}{ |c|c|c|c|c|c| } 
    \hline
    & $R_{\mbox{\tiny{min}}}$ & $R_{\mbox{\tiny{max}}}$ & amplitude & $\xi$ & $\redchisq$ 
    \\ \hline \hline
    eq.~(\ref{shortR}) & $5$ & \rev{$26$} & \rev{$k_s=0.1534(2)$} & \rev{$22.05(6)$} & \rev{$0.92$} \\
    \hline
    eq.~(\ref{largeR2}) & \rev{$6$} & $47$ & \rev{$k_l=0.0415(4)$} & \rev{$22.13(17)$} & \rev{$0.33$} \\
   \hline
    \end{tabular}
    \caption{Results of the fits to the short- and long-distance behaviors, according to the Svetitsky-Yaffe mapping, of the Polyakov loop correlator for $N_t=9$, $\beta=12.15266$ which corresponds to $T/\Tc=0.86$.}
    \label{table:2}
\end{table}

\begin{table}[h!]
\centering
    \begin{tabular}{ |c|c|c|c|c|c|c|c| } 
    \hline
    $N_t$ & & $T/\Tc$ & $R_{\mbox{\tiny{min}}}$ & $R_{\mbox{\tiny{max}}}$ & amplitude & $\xi$ & $\redchisq$ \\ 
    \hline \hline
    \multirow{5}{*}{$6$} & eq.~(\ref{shortR}) & $0.86$ & $5$ & \rev{$16$} & \rev{$k_s=0.1630(4)$} & \rev{$14.13(5)$} & \rev{$1.10$} \\
    & eq.~(\ref{largeR2}) & $0.86$ & $6$ & $47$ & \rev{$k_l=0.0491(5)$} & \rev{$14.25(15)$} & \rev{$0.09$} \\
    \cline{2-8}
    & eq.~(\ref{shortR}) & $0.89$ & $5$ & \rev{$26$} & \rev{$k_s=0.1660(4)$} & \rev{$18.65(7)$} & \rev{$1.30$} \\
    & eq.~(\ref{largeR2}) & $0.89$ & \rev{$6$} & $47$ & \rev{$k_l=0.0473(6)$} & \rev{$18.40(24)$} & \rev{$0.33$} \\
    \hline 
    \multirow{5}{*}{$7$} & eq.~(\ref{shortR}) & $0.83$ & $5$ & \rev{$15$} & \rev{$k_s=0.1562(4)$} & \rev{$13.37(5)$} & \rev{$1.15$} \\
    & eq.~(\ref{largeR2}) & $0.83$ & \rev{$6$} & $47$ & \rev{$k_l=0.04716(24)$} & \rev{$13.62(7)$} & \rev{$0.33$} \\
    \cline{2-8}
    & eq.~(\ref{shortR}) & $0.86$ & $5$ & \rev{$22$} & \rev{$k_s=0.1587(3)$} & \rev{$17.03(5)$} & \rev{$0.83$} \\
    & eq.~(\ref{largeR2}) & $0.86$ & \rev{$6$} & $47$ & \rev{$k_l=0.0458(3)$} & \rev{$17.03(10)$} & \rev{$0.19$} \\
    \hline 
    \multirow{7.5}{*}{$8$} & eq.~(\ref{shortR}) & $0.80$ & $5$ & \rev{$15$} & \rev{$k_s=0.1497(4)$} & \rev{$12.75(5)$} & \rev{$1.55$} \\
    & eq.~(\ref{largeR2}) & $0.80$ & \rev{$6$} & $47$ & \rev{$k_l=0.0454(7)$} & \rev{$13.07(17)$} & \rev{$0.13$} \\
    \cline{2-8}
    & eq.~(\ref{shortR}) & $0.83$ & $5$ & \rev{$20$} & \rev{$k_s=0.1526(4)$} & \rev{$15.65(6)$} & \rev{$1.35$} \\
    & eq.~(\ref{largeR2}) & $0.83$ & $6$ & $47$ & \rev{$k_l=0.0445(4)$} & \rev{$15.86(13)$} & \rev{$0.09$} \\
    \cline{2-8} 
    & eq.~(\ref{shortR}) & $0.86$ & $6$ & \rev{$32$} & \rev{$k_s=0.1543(3)$} & \rev{$20.36(7)$} & \rev{$0.96$} \\
    & eq.~(\ref{largeR2}) & $0.86$ & \rev{$6$} & $47$ & \rev{$k_l=0.0429(4)$} & \rev{$20.05(20)$} & \rev{$0.05$} \\
    \hline 
    \multirow{7.5}{*}{$9$} & eq.~(\ref{shortR}) & $0.80$ & $5$ & \rev{$17$} & \rev{$k_s=0.1462(3)$} & \rev{$14.47(4)$} & \rev{$0.75$} \\
    & eq.~(\ref{largeR2}) & $0.80$ & \rev{$6$} & $47$ & \rev{$k_l=0.0432(4)$} & \rev{$14.74(13)$} & \rev{$0.16$} \\
    \cline{2-8}
    & eq.~(\ref{shortR}) & $0.83$ & $5$ & \rev{$25$} & \rev{$k_s=0.1492(3)$} & \rev{$17.81(7)$} & \rev{$1.01$} \\
    & eq.~(\ref{largeR2}) & $0.83$ & \rev{$6$} & $47$ & \rev{$k_l=0.0422(4)$} & \rev{$17.96(18)$} & \rev{$0.23$} \\
    \cline{2-8} 
    & eq.~(\ref{shortR}) & $0.86$ & $5$ & \rev{$26$} & \rev{$k_s=0.1534(2)$} & \rev{$22.05(6)$} & \rev{$0.92$} \\
    & eq.~(\ref{largeR2}) & $0.86$ & \rev{$6$} & $47$ & \rev{$k_l=0.0415(4)$} & \rev{$22.13(17)$} & \rev{$0.33$} \\
    \hline
    \end{tabular}
    \caption{Results of the fits to the short- and long-distance behavior, according to the Svetitsky-Yaffe mapping, of the Polyakov-loop correlator for different values of $\beta$ and $N_t$.}
    \label{table_confront}
\end{table}

In order to appreciate the agreement, in fig.~\ref{fig_3} we show the results of both fits. 
\vskip 0.3cm



\subsection{Identification of EST corrections beyond the Nambu-Got{\={o}} approximation}

The most efficient way to identify EST corrections beyond the Nambu-Got{\={o}} approximation is to study the behavior of the ground-state energy $E_0$ as a function of the temperature, as the deconfinement transition is approached from below, at a fixed value of $\beta$ (i.e. at fixed lattice spacing $a$) and varying the temperature by changing the value of $N_t$. For this purpose, we performed additional sets of simulations, whose details are reported in tab.~\ref{table:3}, in tab.~\ref{table:4}, and in tab.~\ref{table:4bis}. For each of the configuration ensembles produced in these runs, we extracted the value of $E_0$ (defined as the inverse of the correlation length $\xi$).  
The results of these fits are reported in tab.~\ref{table:5}, in tab.~\ref{table:6}, and in tab.~\ref{table:6bis}, respectively.

\rev{At temperatures close to the deconfinement transition, the large values of the correlation length required a careful treatment of effects due to the finiteness of the spatial extent of the system. We addressed this issue in two ways.
\begin{enumerate}
\item For some $N_t$ values, we repeated our simulations on lattices of larger spatial sizes, $N_s=160$ and $N_s=240$, as shown in tab.~\ref{table:3}, in tab.~\ref{table:4}, and in tab.~\ref{table:4bis}.
\item We generalized eq.~(\ref{largeR2}) by including the contributions to the correlator not only from the first periodic copy of the system, but all of the copies whose contribution was larger than the statistical error of the simulations.
\end{enumerate}
For the remaining sets of simulations, i.e. those corresponding to $N_t\geq 7$ for $\beta=9$, to $N_t\geq 9$ for $\beta=12.15266$, and to $N_t\geq 10$ for $\beta=13.42445$, we fitted our data using eq.~(\ref{largeR2}).}

Studying the EST corrections through the analysis of the ground-state energy has two main advantages. Firstly, in this setting the Nambu-Got{\={o}} expectation for $E_0$ is exactly known and is given by eq.~\ref{E0}, which we recall here
\begin{equation} 
   E_0 = N_t\sigma(N_t)  = N_t\sigma_0\sqrt{1-\frac{\pi}{3N_t^2\sigma_0}} = N_t\sigma_0\sqrt{1-\frac{T^2}{T_{c,NG}^2}}.
   \label{NG}
\end{equation}
Note that eq.~(\ref{NG}) predicts a mean-field critical index $\nu=1/2$ for the correlation length $\xi=1/E_0$: this is obviously incompatible with the prediction from the Svetitsky-Yaffe mapping, from which one would expect the two-dimensional Ising critical index $\nu=1$,
i.e. a linear scaling as $T\to \Tc$ from below:
\begin{equation}
\xi \sim \bigg(1-\frac{T}{\Tc}\bigg)^{-1},
\end{equation}
or, equivalently:
\begin{equation}
\label{eq.5.20}
\xi \sim \bigg(1-\frac{N_{t,c}}{N_t}\bigg)^{-1},
\end{equation}
which translates into the following \emph{Ansatz} for the form of the ground-state energy in the vicinity of the deconfinement point:
\begin{equation}
\label{linear}
    E_0 \sim 1-\frac{N_{t,c}}{N_t}.
\end{equation}
Thus, the dependence of $E_0$ on the temperature is an ideal probe for corrections with respect to the Nambu-Got{\={o}} approximation.

Moreover, this approach follows very closely the one used in the bootstrap analysis of ref.~\cite{EliasMiro:2019kyf,Miro:2021rof}. From those works, using the low-energy universality we can search for numerical evidence of the first correction to the Nambu-Got{\={o}} approximation, which is expected to appear at the order $1/N_t^7$ and can be parametrized as
\begin{equation}
E_0(N_t)= N_t\sigma_0\sqrt{1-\frac{\pi}{3N_t^2\sigma_0}} - \frac{32 \pi^6 \gamma_3}{225 \sigma_0^3 N_t^7}.
\end{equation}

\begin{table}[h]
\centering
    \begin{tabular}{ |c|c|c|c|c| } 
    \hline
    $\beta$ & $N_t$ & $N_s$ & $T/\Tc$ & $\nconf$ \\ 
    \hline \hline
    \multirow{10.55}{*}{$9$} & $6$ & \rev{$160$} & $0.935$ & $2.0\times10^5$ \\
    \cline{2-4}
    & $7$ & $96$ &$0.801$ & $2.0\times10^5$ \\
    \cline{2-4}
    & $8$ & $96$ &$0.701$ & $2.0\times10^5$ \\ 
    \cline{2-4}
    & $9$ & $96$ &$0.623$ & $2.0\times10^5$ \\
    \cline{2-4}
    & $10$ & $96$ &$0.561$ & $2.0\times10^5$ \\
    \cline{2-4}
    & $11$ & $96$ &$0.510$ & $2.0\times10^5$ \\
    \cline{2-4}
    & $12$ & $96$ &$0.468$ & $2.0\times10^5$ \\
    \hline
    \end{tabular}
    \caption{Information on the simulations at $\beta=9$.}
    \label{table:3}
\end{table}

\begin{table}[h]
\centering
    \begin{tabular}{ |c|c|c|c|c| } 
    \hline
    $\beta$ & $N_t$ &  $N_s$ &$T/\Tc$ & $\nconf$ \\ 
    \hline \hline
    \multirow{10.57}{*}{$12.15266$} & $8$ &\rev{$240$} & $0.960$ & $2.0\times10^5$ \\
    \cline{2-4}
    & $9$ & \rev{$160$} &$0.853$ & $2.0\times10^5$ \\
    \cline{2-4}
    & $10$ & $96$ & $0.768$ & $2.0\times10^5$ \\ 
    \cline{2-4}
    & $11$ & $96$ &$0.698$ & $2.0\times10^5$ \\
    \cline{2-4}
    & $12$ & $96$ &$0.640$ & $2.0\times10^5$ \\
    \cline{2-4}
    & $13$ & $96$ &$0.591$ & $2.0\times10^5$ \\
    \cline{2-4}
    & $14$ & $96$ &$0.549$ & $2.0\times10^5$ \\
    \hline
    \end{tabular}
    \caption{Information on the simulations at $\beta=12.15266$.}
    \label{table:4}
\end{table}

\begin{table}[h]
\centering
    \begin{tabular}{ |c|c|c|c|c| }
    \hline
    $\beta$ & $N_t$ &  $N_s$ & $T/\Tc$ & $\nconf$ \\
    \hline \hline
    \multirow{10.57}{*}{$13.42445$} & $9$ &\rev{$240$} & $0.947$ & $2.0\times10^5$ \\
    \cline{2-4}
    & $10$ & \rev{$160$} &$0.852$ & $2.0\times10^5$ \\
    \cline{2-4}
    & $11$ & \rev{$160$} &$0.775$ & $2.0\times10^5$ \\
    \cline{2-4}
    & $12$ & $96$ &$0.710$ & $2.0\times10^5$ \\
    \cline{2-4}
    & $13$ & $96$ &$0.655$ & $2.0\times10^5$ \\
    \cline{2-4}
    & $14$ & $96$ &$0.609$ & $2.0\times10^5$ \\
    \cline{2-4}
    & $15$ & $96$ &$0.568$ & $2.0\times10^5$ \\
    \hline
    \end{tabular}
    \caption{Information on the simulations at $\beta=13.42445$.}
    \label{table:4bis}
\end{table}

\begin{table}[h]
\centering
  \begin{tabular}{ |c|c|c|c|c|c| } 
 \hline
 $N_t$ & $R_{\mbox{\tiny{min}}}$ & $R_{\mbox{\tiny{max}}}$ & $k_l$ & $E_0$ & $\redchisq$
    \\ \hline \hline
    $6$ & \rev{$7$} & $47$ & \rev{$0.0417(6)$} & \rev{$0.0294(5)$} & \rev{$0.07$} \\
    \hline
    $7$ & \rev{$7$} & $47$ & \rev{$0.0495(9)$} & \rev{$0.0871(13)$} & \rev{$0.58$} \\
    \hline
    $8$ & \rev{$8$} & $47$ & \rev{$0.0475(15)$} & \rev{$0.1250(25)$} & \rev{$0.56$} \\
    \hline
    $9$ & \rev{$8$} & $47$ & \rev{$0.0501(15)$} & \rev{$0.1691(36)$} & \rev{$0.30$} \\
    \hline
    $10$ & \rev{$9$} & $47$ & \rev{$0.0432(33)$} & \rev{$0.1941(63)$} & \rev{$2.00$} \\
    \hline
    $11$ & \rev{$9$} & $47$ & \rev{$0.0469(41)$} & \rev{$0.2381(80)$} & \rev{$0.72$} \\
    \hline
    $12$ & \rev{$9$} & $47$ & \rev{$0.0392(32)$} & \rev{$0.2581(69)$} & \rev{$0.09$} \\
    \hline
    \end{tabular}
    \caption{Best-fit estimates for $E_0$ for different values of $N_t$ at $\beta=9$.}
    \label{table:5}
\end{table}

\begin{table}[h]
\centering
    \begin{tabular}{ |c|c|c|c|c|c| } 
    \hline
    $N_t$ & $R_{\mbox{\tiny{min}}}$ & $R_{\mbox{\tiny{max}}}$ & $k_l$ & $E_0$ & $\redchisq$ 
    \\ \hline \hline
    $8$ & \rev{$8$} & $47$ & \rev{$0.0338(10)$} & \rev{$0.0135(5)$} & \rev{$0.36$} \\
    \hline
    $9$ & \rev{$10$} & $47$ & \rev{$0.0416(4)$} & \rev{$0.0452(3)$} & \rev{$0.33$} \\
    \hline
    $10$ & \rev{$9$} & $47$ & \rev{$0.0416(13)$} & \rev{$0.0695(17)$} & \rev{$0.45$} \\
    \hline
    $11$ & \rev{$10$} & $47$ & \rev{$0.0423(16)$} & \rev{$0.0922(22)$} & \rev{$0.35$} \\
    \hline
    $12$ & \rev{$10$} & $47$ & \rev{$0.0414(20)$} & \rev{$0.1132(32)$} & \rev{$0.13$} \\
    \hline
    $13$ & \rev{$11$} & $47$ & \rev{$0.0408(26)$} & \rev{$0.1334(42)$} & \rev{$0.25$} \\
    \hline
    $14$ & \rev{$11$} & $47$ & \rev{$0.0355(16)$} & \rev{$0.1440(30)$} & \rev{$0.18$} \\
    \hline
    \end{tabular}
    \caption{Best-fit estimates for $E_0$ for different values of $N_t$ at $\beta=12.15266$.}
    \label{table:6}
\end{table}

\begin{table}[h]
\centering
    \begin{tabular}{ |c|c|c|c|c|c| }
    \hline
    $N_t$ & $R_{\mbox{\tiny{min}}}$ & $R_{\mbox{\tiny{max}}}$ & $k_l$ & $E_0$ & $\redchisq$
    \\ \hline \hline
    $9$ & \rev{$9$} & $47$ & \rev{$0.0338(13)$} & \rev{$0.0151(7)$} & \rev{$0.32$} \\
    \hline
    $10$ & \rev{$11$} & $47$ & \rev{$0.0391(6)$} & \rev{$0.0401(5)$} & \rev{$0.06$} \\
    \hline
    $11$ & \rev{$12$} & $47$ & \rev{$0.0407(8)$} & \rev{$0.0619(8)$} & \rev{$0.05$} \\
    \hline
    $12$ & \rev{$11$} & $47$ & \rev{$0.0387(7)$} & \rev{$0.0777(9)$} & \rev{$0.52$} \\
    \hline
    $13$ & \rev{$12$} & $47$ & \rev{$0.0399(12)$} & \rev{$0.0979(16)$} & \rev{$0.06$} \\
    \hline
    $14$ & \rev{$12$} & $47$ & \rev{$0.0390(13)$} & \rev{$0.1135(20)$} & \rev{$0.19$} \\
    \hline
    $15$ & \rev{$12$} & $47$ & \rev{$0.0359(17)$} & \rev{$0.1254(27)$} & \rev{$0.25$} \\
    \hline
    \end{tabular}
    \caption{Best-fit estimates for $E_0$ for different values of $N_t$ at $\beta=13.42445$.}
    \label{table:6bis}
\end{table}

\begin{figure}[!htb]
  \centering
  \includegraphics[width=0.95\textwidth, clip]{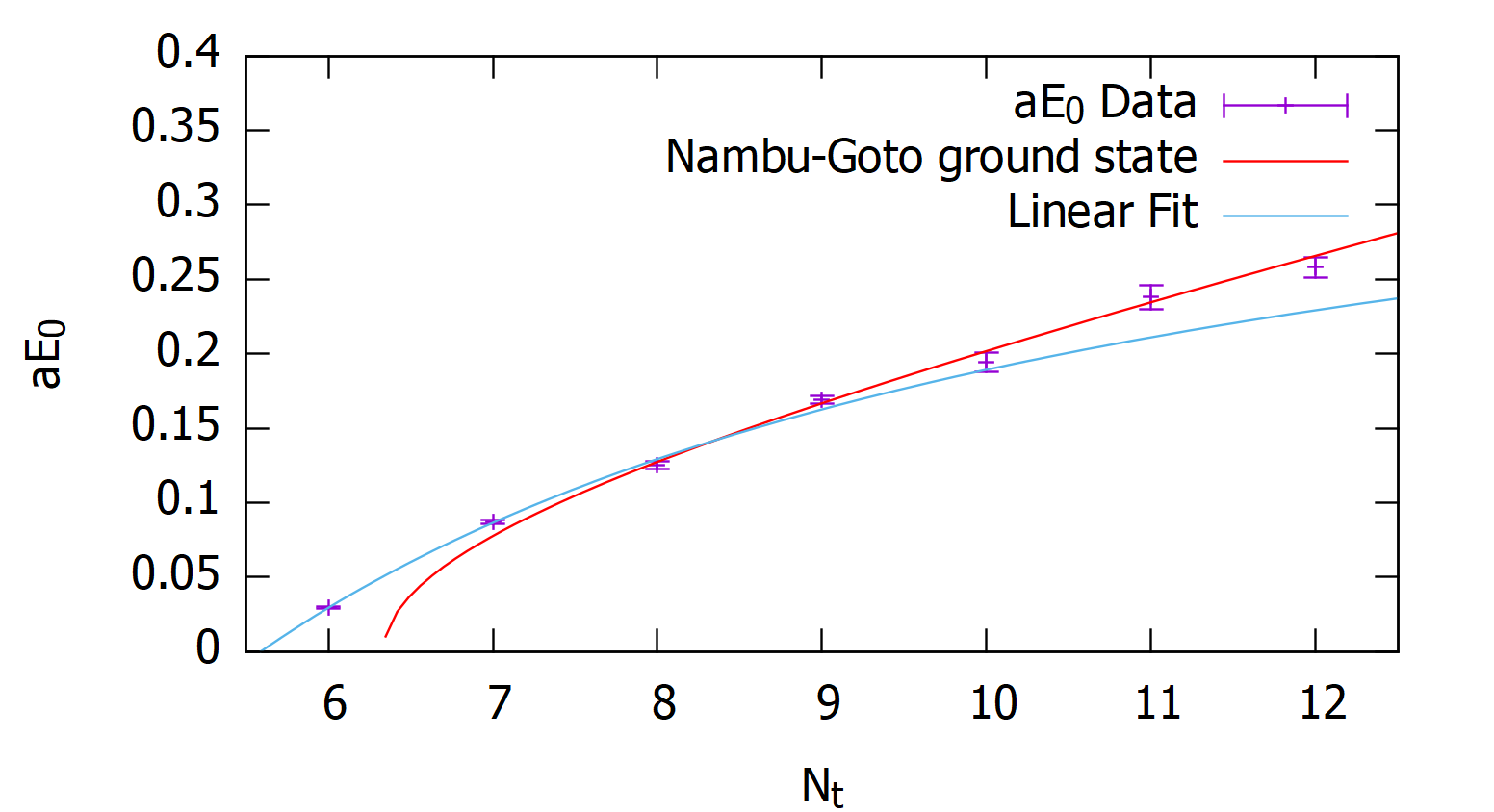}
  \caption{Fits of our data for the ground-state energy $E_0$ at $\beta=9$ according to the prediction from the Nambu-Got{\={o}} model, eq.~(\ref{NG}) (red line) and according to the \emph{Ansatz} of a linear dependence of $E_0$ on the temperature, according to eq.~(\ref{linear}) (blue curve). Note that the quantity on the horizontal axis of this plot is the inverse of the temperature, in units of the lattice spacing.}\label{fig_5}
\end{figure}

\begin{figure}[!htb]
  \centering
  \includegraphics[width=0.95\textwidth, clip]{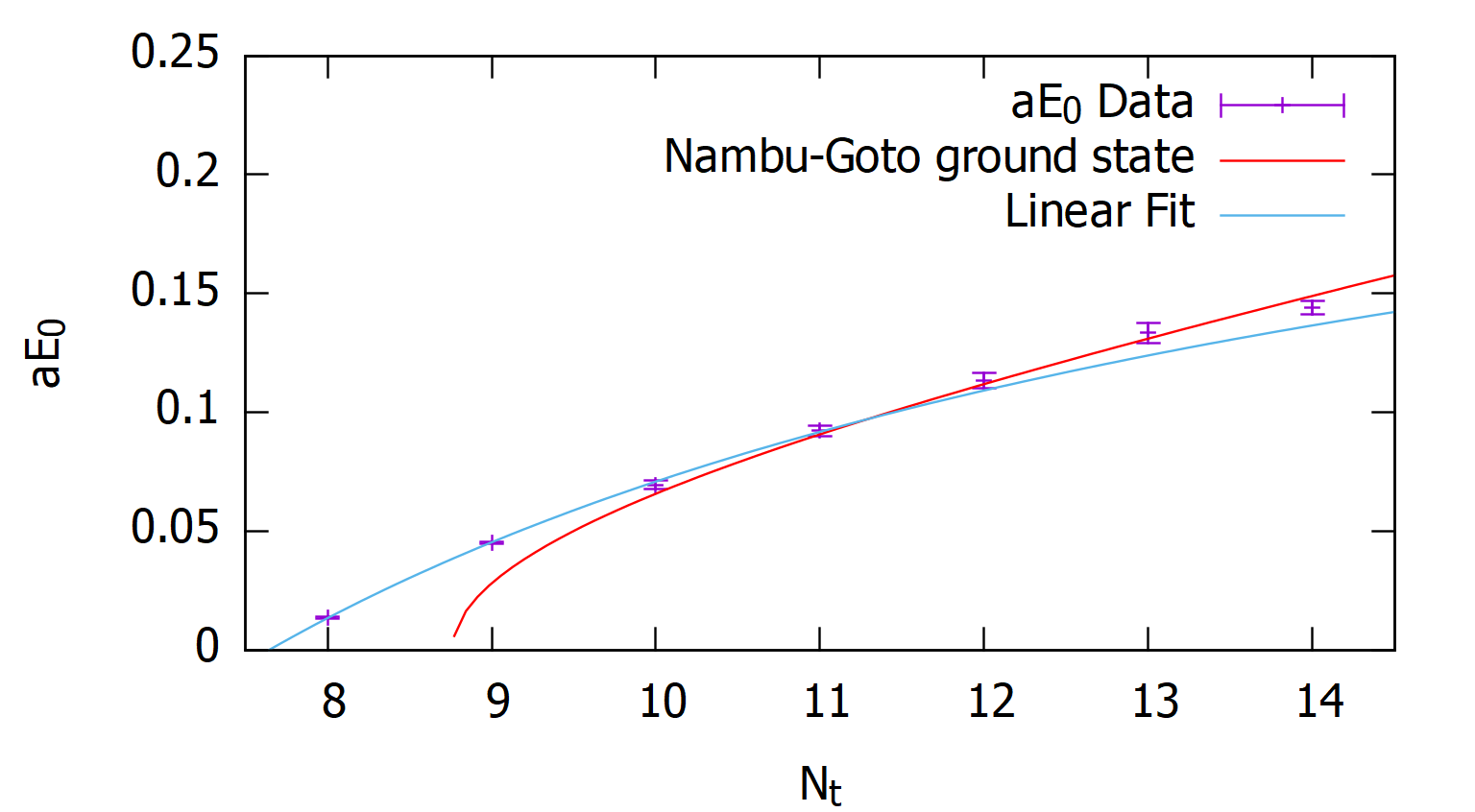}
  \caption{Same as in fig.~\ref{fig_5}, but for the data at $\beta=12.15266$.}\label{fig_6}
\end{figure}

\begin{figure}[!htb]
  \centering
  \includegraphics[width=0.95\textwidth, clip]{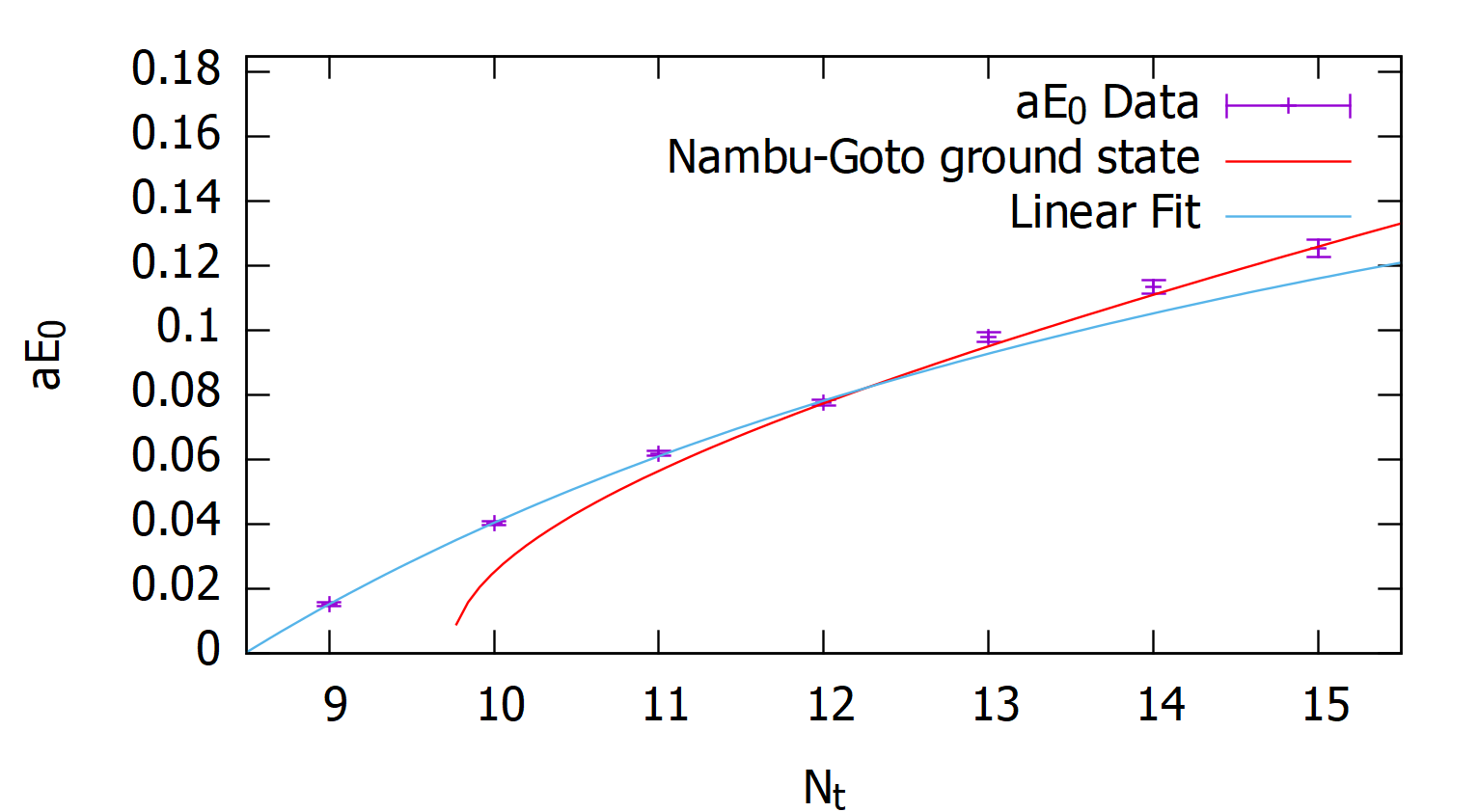}
  \caption{Same as in fig.~\ref{fig_5}, but for $\beta=13.42445$.}\label{fig_6a}
\end{figure}

Following these observations, we first tried to fit the data using a functional form motivated by the Nambu-Got{\={o}} model ($\nu=1/2$) and one based on the Ising model ($\nu=1$). Unsurprisingly, we found that neither choice describes the data accurately: this is clearly visible in fig.~\ref{fig_5}, in fig.~\ref{fig_6}, and in fig.~\ref{fig_6a}. For all the three $\beta$ values the Nambu-Got{\={o}} curve fits the data well at large $N_t$ (i.e. at low temperature), but it misses the approach to the critical point, when the deconfinement transition is approached. On the contrary, the linear fit agrees with the data near the critical point, as expected from the Svetitsky-Yaffe correspondence, but this agreement holds only for the first few values of $N_t$. For larger values, a linear fit is not consistent with the Monte~Carlo data.

This failure is indeed in agreement with the low-energy universality: the latter suggests that the correct behavior at short distances should not be modelled by assuming an expression like the one in eq.~(\ref{linear}), but rather by adding a suitable $1/N_t^7$ correction to the Nambu-Got{\={o}} approximation.

Moreover, since low-energy universality suggests that, starting from $O(N_t^{-7})$, there may also be any possible higher-order corrections, we truncate for consistency the Nambu-Got{\={o}} expression to this order. Based on this reasoning, we assume the following form for the $N_t$ dependence of the ground state energy:
\begin{equation}
\label{universality}
E_0(N_t)=\mbox{Taylor}_4(E_0)+\frac{k_{4}}{(\sigma_0)^3N_t^7},
\end{equation}
with
\begin{equation} 
 \mbox{Taylor}_4(E_0) \equiv  \sigma_0 N_t - \frac{\pi}{6N_t}-\frac{\pi^2}{72(\sigma_0)N_t^3}-\frac{\pi^3}{432(\sigma_0)^2N_t^5} -\frac{5\pi^4}{10368(\sigma_0)^3N_t^7},
\label{Taylor}
\end{equation}
where the Taylor expansion is completely known, and the only free parameters of the fit are the zero-temperature string tension $\sigma_0$ and the $k_4$ coefficient.

\begin{table}[h]
\centering
    \begin{tabular}{ |c|c|c|c|c|c|c| }
    \hline
    $\beta$ & $N_{t,\mbox{\tiny{min}}}$ & $N_{t,\mbox{\tiny{max}}}$ & $k_4$ & $\sigma_0$ & $\redchisq$ & literature 
    \\ \hline \hline
    $9$ & $6$ & $12$ & \rev{$0.040(8)$} & \rev{$0.02603(19)$} & \rev{$1.60$} & $0.02583(3)$ \\
    \hline
    $12.15266$ & $8$ & $14$ & \rev{$0.054(5)$} & \rev{$0.01366(5)$} & \rev{$0.89$} & $0.01371(29)$\\
    \hline
    $13.42445$ & $9$ & $15$ & \rev{$0.053(8)$} & \rev{$0.01104(5)$} & \rev{$1.33$} & $0.01108(23)$ \\
    \hline
    \end{tabular}
    \caption{Results of the fits of our numerical data to eq.~(\ref{universality}). In the last column we report the values of $\sigma_0$ quoted in ref.~\cite{Bonati:2021vbc} for $\beta=9$, and in ref.~\cite{Teper:1998te} for $\beta=12.15266$ and for $\beta=13.42445$.}
    \label{table:7}
\end{table}

\begin{table}[h]
\centering
    \begin{tabular}{ |c|c|c|c|c| }
    \hline
    $\beta$ & $N_{t,\mbox{\tiny{min}}}$ & $N_{t,\mbox{\tiny{max}}}$ & $k_4$ & $\redchisq$ 
    \\ \hline \hline
    $9$ & $6$ & $12$ & \rev{$0.048(3)$} & \rev{$1.62$}  \\
    \hline
    $12.15266$ & $8$ & $14$ & \rev{$0.049(2)$}  & \rev{$0.89$} \\
    \hline
    $13.42445$ & $9$ & $15$ & \rev{$0.048(4)$}  & \rev{$1.24$} \\
    \hline
    \end{tabular}
    \caption{\rev{Results of the fits of our numerical data to eq.~(\ref{universality}) using as input the values for $\sigma_0$ obtained from the literature.}}
    \label{table:7bis}
\end{table}

\begin{figure}[!htb]
  \centering
  \includegraphics[width=0.95\textwidth, clip]{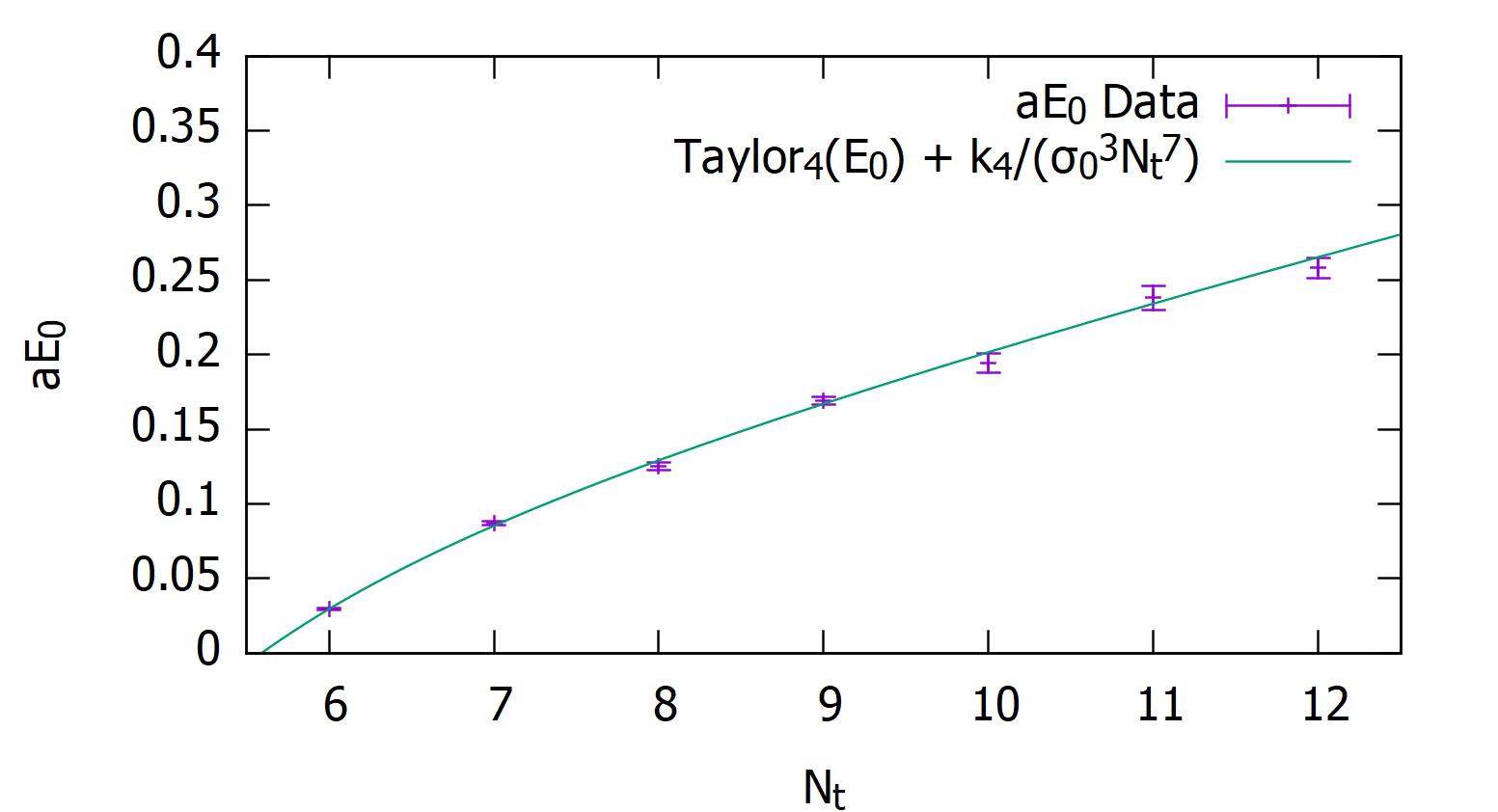}
  \caption{Fit of our numerical results for the ground-state energy at $\beta=9$ to eq.~(\ref{universality}).}\label{fig_7a}
\end{figure}
 
\begin{figure}[!htb]
  \centering
  \includegraphics[width=0.95\textwidth, clip]{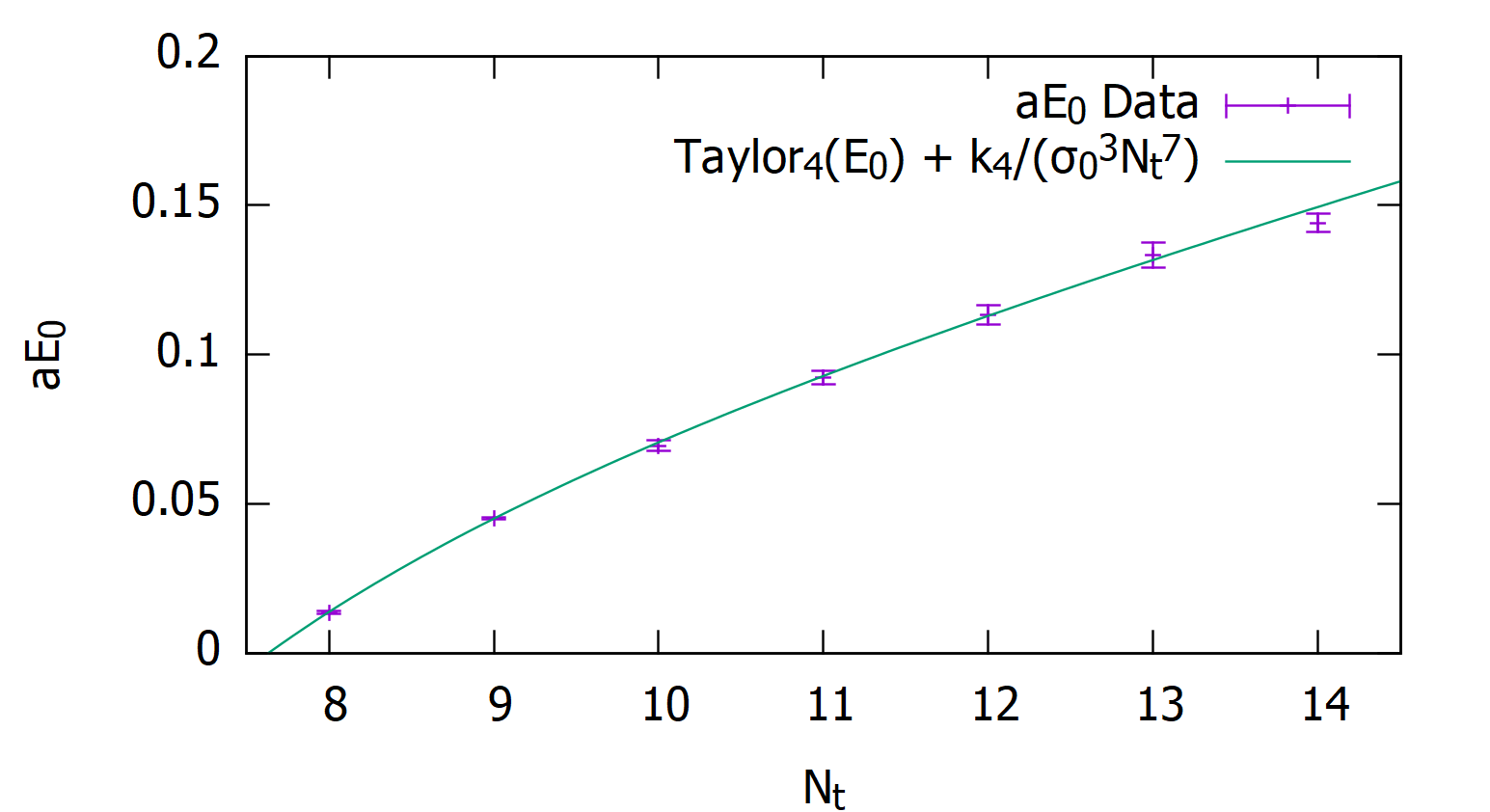}
  \caption{Same as in fig.~\ref{fig_7a}, but for $\beta=12.15266$.}\label{fig_7b}
\end{figure}
 
\begin{figure}[!htb]
  \centering
  \includegraphics[width=0.95\textwidth, clip]{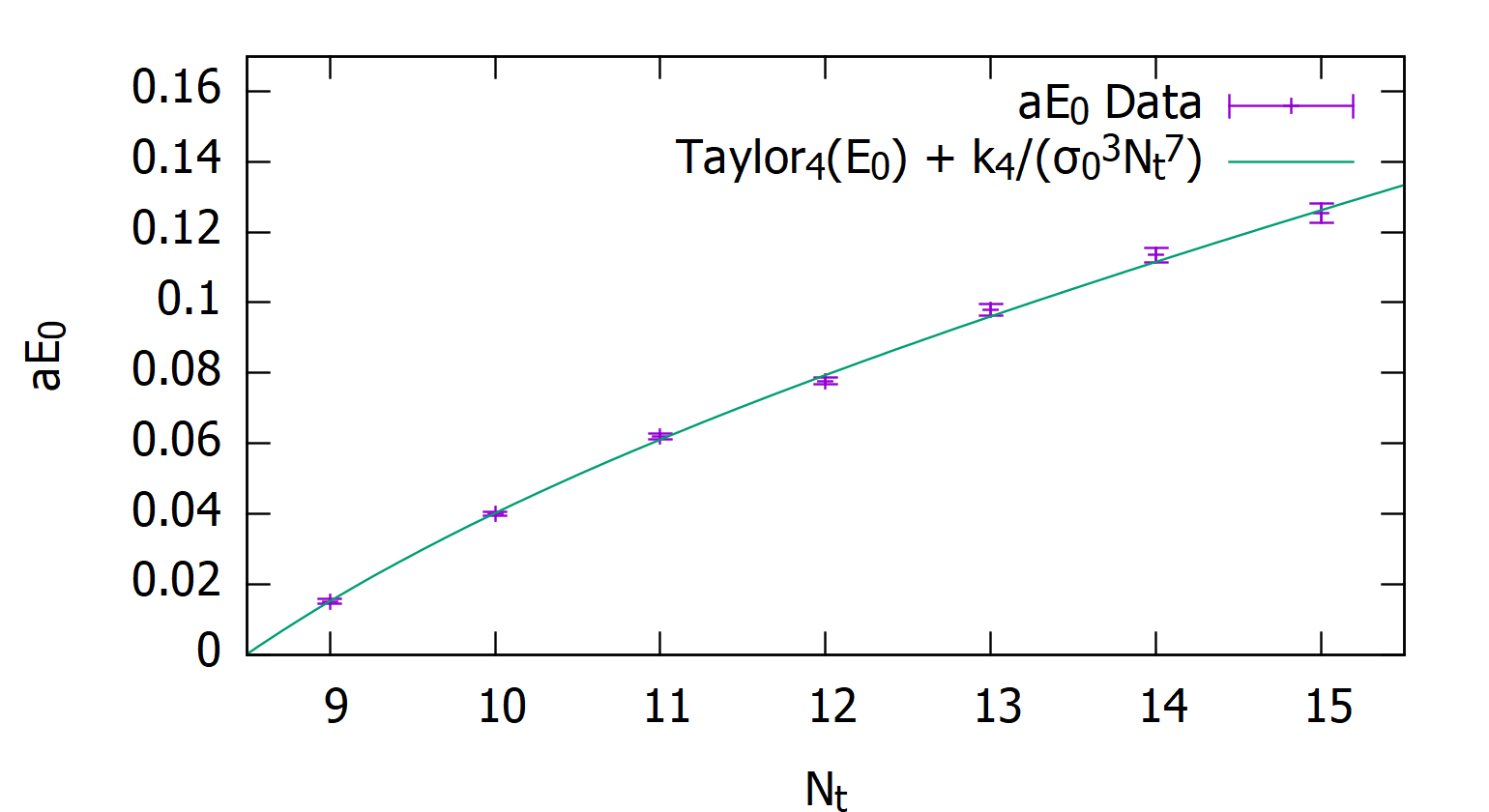}
  \caption{Same as in fig.~\ref{fig_7a}, but for $\beta=13.42445$.}\label{fig_7c}
\end{figure}

Remarkably, these fits yield very good reduced $\chi^2$ values for the data at all the three $\beta$ values. The detailed results are reported in tab.~\ref{table:7}, and shown in fig.~\ref{fig_7a}, in fig.~\ref{fig_7b} and in fig.~\ref{fig_7c}. In particular, the quality of the fits improves as one goes toward the continuum limit. Moreover, the best-fit values obtained for $\sigma_0$ are fully consistent with those that were independently obtained in ref.~\cite{Bonati:2021vbc} (for $\beta=9$) and in ref.~\cite{Teper:1998te} (for $\beta=12.15266$ and for $\beta=13.42445$). In fact, the precision of our results for $\sigma_0$ is even better than the one from the latter reference, so that, in principle, the approach that we used for the determination of $\sigma_0$ could even be used for scale setting (although much more precise scale-setting methods exist today~\cite{Luscher:2010iy, Francis:2015lha}). 

Another non-trivial consistency check of our analysis is that the three values of $k_4$ that we found should agree within their uncertainties, since the scale dependence of this coefficient is already accounted for by the $1/\sigma_0^3$ normalization in eq.~(\ref{universality}). As tab.~\ref{table:7} shows, this expectation is indeed confirmed in our fit results, and the three values are compatible within their errors.
\rev{This agreement is also confirmed by the analysis reported in tab.~\ref{table:7bis}, where we performed the same fits as in tab.~\ref{table:7} but fixing $\sigma_0$ to the values reported in the literature. The agreement between the three values of $\beta$ that we tested is clear. We remark that all of these data could be fitted with only one free parameter, $k_4$.} 

\rev{As our final result for $k_4$, we quote a weighted average of the three values from tab.~\ref{table:7}: $k_4=0.050(8)$, with a somewhat conservative estimate for the uncertainty. From this result, using eq.~(\ref{gamma3}), we obtain
\begin{equation}
\gamma_3=-\frac{225}{32\pi^6} k_4 = \rev{- 0.00037(6)},
\end{equation}
}
which is well inside the bound $\gamma_3\geq-\frac{1}{768}\simeq -0.0013$ derived in refs.~\cite{EliasMiro:2019kyf,Miro:2021rof}.

The fact that $\gamma_3$ is negative \rev{was already noted in ref.~\cite{Athenodorou:2016kpd}} and is non-trivial: in particular, \rev{as shown in ref.~\cite{EliasMiro:2019kyf}, it does not allow to prove the Axionic String \emph{Ansatz} for the EST describing this gauge theory.}

\rev{The contribution associated with the $\gamma_3$ term is only the first term of an infinite series of higher-order corrections to the Nambu-Got{\={o}} action. Unfortunately, our data do not allow us to extract information about the subleading terms, due to the fact that discretization effects on $N_t$ are too large. All our attempts to fit the data adding a further term in the Taylor expansion, i.e. fitting the data with: 
\begin{equation}
\label{universality2}
E_0(N_t)=\mbox{Taylor}_5(E_0)+\frac{k_{4}}{(\sigma_0)^3N_t^7}+\frac{k_{5}}{(\sigma_0)^4N_t^9},
\end{equation}
led to values for $k_4$ and $k_5$ compatible with zero within their uncertainties (the two terms tend to compensate each other in the fit).

We can compare our results with those obtained in ref.~\cite{Athenodorou:2016kpd} at $\beta=16.0$. Fitting the values for $E_0$ reported in ref.~\cite{Athenodorou:2016kpd} with eq.~(\ref{universality}), one finds $k_4\simeq 0.80$, but using instead eq.~(\ref{universality2}) one finds $k_4=0.069(25)$ and $k_5=0.045(22)$, with a string tension $\sigma_0=0.007644(4)$, in full agreement with the value quoted in ref.~\cite{Athenodorou:2016kpd}, which is $\sigma_0=0.0076416(46)$, and a reduced $\chi^2=1.33$. This $k_4$ value is compatible with the one that we found at smaller values of $\beta$, suggesting that simulations at even larger $\beta$ may lead to reliable estimates also for $k_5$.}

Finally, it is interesting to compare our result for $\gamma_3$ with the one obtained in refs.~\cite{Dubovsky:2014fma,Conkey:2016qju,Chen:2018keo} for the $\SU(6)$ Yang-Mills theory in three dimensions, using the lattice data from ref.~\cite{Athenodorou:2011rx}, which was of similar magnitude but opposite in sign. \rev{Moreover, estimates for $k_4$ can be extracted for the $\SU(4)$, $\SU(6)$, and $\SU(8)$ theories using the results for the ground-state energy from ref.~\cite{Athenodorou:2016kpd}: this leads to results that are different for the different gauge groups, showing} explicitly that, at this level of resolution, the EST is not universal anymore: instead it encodes, as it should, the specific properties of the underlying Yang-Mills theory.

\section{Concluding remarks}
\label{sec:concluding_remarks}

In this work we discussed the results of a set of high-precision simulations of the Polyakov-loop correlator in the $\SU(2)$ lattice gauge theory in three dimensions. All our simulations were run at finite temperature, in the vicinity of the deconfinement transition, in the range 
$0.8\Tc\leq T \leq \Tc$, where it is easier to compare the simulations with the EST predictions. Moreover, invoking the Svetitsky-Yaffe conjecture, in this regime one can compare the analytical solution of the two-dimensional Ising model with the gauge-theory data. The results of this comparison revealed remarkable agreement between our numerical results for the Polyakov-loop correlator and the exact expression of spin-spin correlator of the two-dimensional Ising model. We could extract very precise values for the ground-state energy $E_0$ of the effective string describing this gauge theory, and quantify the deviations from the predictions that can be derived approximating the effective string action with the Nambu-Got{\={o}} action.
 
We conclude with some comments on these results.
\begin{enumerate}
\item The type of comparison that we carried out is not limited to theories having a critical point in the universality class of the two-dimensional Ising model (or to another exactly integrable model). Indeed, in principle conformal perturbation theory allows one to work out the form of the spin-spin correlator for any model characterized by a continuous phase transition. As a consequence, the Svetitsky-Yaffe mapping can be used also for spin models that are not exactly integrable. An example of this approach was discussed in ref.~\cite{Caselle:2019tiv}, where results for the $\SU(2)$ Yang-Mills theory in four dimensions were compared with the Ising model in three dimensions.
\item As we discussed in section~\ref{sec:est}, an interesting consequence of the effective string description is that the long-distance behavior of the correlator of a Yang-Mills theory in $(d+1)$ dimensions is dominated by a $K_{(d-2)/2}(E_0)$ Bessel function, in exact agreement with the long-distance behavior predicted by the Svetitsky-Yaffe conjecture for the spin-spin correlator of the underlying $d$-dimensional spin model. This holds for any EST (under very mild conditions) and for any spin model, and can be regarded as a non-trivial check of mutual consistency for the effective string and the Svetitsky-Yaffe conjecture.
\item The value that we found for the $\gamma_3$ coefficient for the EST describing the $\SU(2)$ Yang-Mills theory in three dimensions \rev{does not allow to prove the Axionic String \emph{Ansatz}}. It would be interesting to extend our analysis to the $\SU(4)$ gauge theory, for which results compatible with the Axionic String \emph{Ansatz} were recently obtained in refs.~\cite{Dubovsky:2016cog,Conkey:2019blu}. Similarly, it would also be interesting to explore the same type of contribution to the EST describing gauge theories in four dimensions, as the axionic string is expected to play an important role in the description of the low-energy dynamics of quantum chromodynamics~\cite{Dubovsky:2015zey}.
\item Historically, one of the problems of the EST description of Yang-Mills theories was its universality, i.e. the fact that it predicted essentially the same behavior (with only a mild dependence on the number of spacetime dimensions), for models as different as the three-dimensional $\Z_2$ gauge model as the four-dimensional $\SU(3)$ Yang-Mills theory. This feature is now understood as a universality that manifests itself only at low energy (or, equivalently, a side-effect of the high accuracy of the Nambu-Got{\={o}} approximation of EST), while the details related to the gauge group (and, possibly, to the confining mechanism into play) may be encoded in the higher-order EST corrections, which are not expected to be universal. In view of the fact that our results indicate that the three-dimensional $\SU(2)$ gauge theory \rev{is not necessarily  described} by an axionic effective string, while the opposite conclusion was recently obtained for the $\SU(6)$ Yang-Mills theory~\cite{Dubovsky:2014fma,Conkey:2016qju,Chen:2018keo}, quantifying these corrections and understanding how they depend on the gauge group would be of great importance. The numerical precision of current Monte~Carlo studies of lattice gauge theory is sufficient to probe the fine details of the effective string theory at very high orders, and to test the accurate theoretical predictions that have been formulated during the past few years~\cite{Brandt:2016xsp}. This is possible thanks not only to the increase in computing power, but also to the deployment of increasingly sophisticated simulation algorithms, among which we would like to mention non-equilibrium Monte~Carlo calculations~\cite{Caselle:2018kap, Francesconi:2020fgi} based on Jarzynski's theorem~\cite{Jarzynski:1996oqb, Jarzynski:1997ef}: as an example of their use for the problems discussed in the present paper, we refer the readers to the recent study of terms $O(L^{-7})$ in the effective-string description of a fluctuating interface of linear size $L$~\cite{Caselle:2016wsw}.
\end{enumerate}

\vskip1.0cm 
\noindent{\bf Acknowledgements}\\

The numerical simulations were run on machines of the Consorzio Interuniversitario per il Calcolo Automatico dell'Italia Nord Orientale (CINECA). We acknowledge support from the SFT Scientific Initiative of INFN.

\bibliography{paper.bib}

\end{document}